\documentclass[reprent,aps,prb,twocolumn,
superscriptaddress,nofootinbib]{revtex4-2}
\usepackage{amsmath,amssymb}
\usepackage{graphicx}
\usepackage{bm}
\usepackage{multirow}
\usepackage{hyperref}
\usepackage{braket}

\usepackage{color}

\begin{document}

\title{
Symmetry analysis of light-induced magnetic interactions via Floquet engineering}
\author{Ryota Yambe}
\email{yambe@g.ecc.u-tokyo.ac.jp}
\affiliation{Department of Applied Physics, The University of Tokyo, Tokyo 113-8656, Japan }
\author{Satoru Hayami}
\email{hayami@phys.sci.hokudai.ac.jp}
\affiliation{Graduate School of Science, Hokkaido University, Sapporo 060-0810, Japan}

\begin{abstract}
Anisotropic magnetic interactions become the origins of intriguing magnetic structures, such as helical and skyrmion structures by the Dzyaloshinskii-Moriya interaction. 
In general, possible anisotropic exchange interactions are restricted by crystal symmetry.
Meanwhile, by lowering the crystal symmetry with light, additional anisotropic magnetic interactions are expected according to its polarization and frequency.
In this study, we clarify a relationship between anisotropic magnetic interactions and symmetry lowering in insulating magnets irradiated by light.
Based on the Floquet formalism, we find that a variety of anisotropic two-spin and three-spin interactions are induced via spin-dependent electric polarizations activated by light irrespective of the presence/absence of the spatial inversion symmetry; we systematically classify them in the hexagonal point group, tetragonal point group, and their subgroups.
Our symmetry analyses show that the light-induced two-spin (three-spin) interaction is owing to the reduction of the point group to a chiral point group (black and white magnetic point group).
We also demonstrate the effect of the light-induced magnetic interactions on the magnetic structures in a triangular unit.
Our results will be a symmetry-based reference for the Floquet engineering of magnetic structures.  
\end{abstract}

\maketitle

\section{Introduction}

Floquet engineering of physical properties has attracted much attention in various fields of condensed matter physics, which gives us a framework to understand the time evolution driven by a time-periodic field within a time-independent Hamiltonian~\cite{eckardt2017colloquium,oka2019floquet,rudner2020floquet}.
In magnetic systems, the effect of such a time-periodic field appears in the modification of magnetic interactions, which results in a variety of phase transitions and associated material design.  
For example, a periodic electric field to the Mott insulator brings about the change of the sign and amplitude of exchange interactions~\cite{mentink2015ultrafast,mentink2017manipulating,PhysRevLett.121.107201,PhysRevB.99.205111,bukov2016schrieffer,chaudhary2019orbital,PhysRevLett.126.177201} and the induction of multiple-spin interactions~\cite{PhysRevB.96.014406,claassen2017dynamical,quito2021polarization} depending on the intensity, polarization, and frequency of the light.
Moreover, combined with the effect of the spin-orbit coupling, the control of magnetic anisotropic exchange interactions by light is possible, as demonstrated in the noncentrosymmetric magnet~\cite{losada2019ultrafast} and the Kitaev magnet~\cite{PhysRevB.103.L100408,PhysRevB.104.214413,PhysRevB.105.085144,kumar2022floquet,PhysRevResearch.4.L032036}.

In general, possible anisotropic exchange interactions are restricted by the crystal symmetry.
For example, the Dzyaloshinskii-Moriya (DM) interaction~\cite{dzyaloshinsky1958thermodynamic,moriya1960anisotropic} appears only when the space inversion symmetry in the lattice structure is absent.
Thus, it is difficult to control the type of anisotropic exchange interactions once the crystal symmetry is determined.
Meanwhile, by introducing the external field, one can expect various symmetry lowerings depending on the direction and polarization of the light, which gives rise to additional anisotropic exchange interactions that are prohibited in the underlying lattice structure. 
This leads to the possibility of engineering any magnetic structures with desired functionalities, which will provide a new guideline for Floquet engineering of the magnetic structure.

This paper gives a symmetry-based understanding of light-induced anisotropic magnetic interactions. 
Based on the Floquet formalism in previous studies~\cite{sato2014floquet,PhysRevLett.117.147202}, we show that a spin-dependent electric polarization activated by light can be the origin of various types of anisotropic magnetic interactions, such as an anisotropic two-site two-spin interaction, anisotropic two-site three-spin interaction, and anisotropic three-site three-spin interaction, by performing group theory and perturbation analyses. 
We classify the light-induced magnetic interactions for the hexagonal point group, tetragonal point group, and their subgroups in a systematic manner.
As a result, we show a comprehensive correspondence between the 
emergent magnetic interactions and symmetry lowering by light. 
We also demonstrate that the light-induced magnetic interactions favor a noncoplanar spin structure with the spin scalar chirality on a triangular unit even without an external static magnetic field.
Our symmetry analysis of the light-induced magnetic interactions will be a reference for controlling magnetic structures by light.

The rest of this paper is organized as follows.
We introduce a static model and time-dependent light-driven model in Secs.~\ref{sec:model_static} and \ref{sec:model_time}, respectively.
We show that the time-dependent model is mapped onto the effective time-independent magnetic interactions by using Floquet theory in Sec.~\ref{sec:model_effective}. 
We present symmetry rules for the light-induced two-site two-spin interaction, two-site three-spin interaction, and three-site three-spin interaction under crystallographic point groups in Secs.~\ref{sec:2spin}-\ref{sec:3site}.
We apply the result to a system consisting of a triangular unit in Sec.~\ref{sec:application}.    
We summarize the paper in Sec.~\ref{sec:summary}.

\section{Model}
\label{sec:model}

\begin{table}
\caption{
\label{tab:class_static}
Symmetry analysis of $J_{12}$, $Y^x_{12}$, and $Y^y_{12}$ for the $\langle 1,2\rangle$ bond along the $x$ direction on the $xy$ plane under 15 point groups.
In the point group \textbf{G}, the 1st, 2nd, and 3rd axes are $x$, $y$, and $z$, respectively.
$\parallel\sharp$ ($\perp\sharp$) for $\sharp=\hat{\bm{x}},\hat{\bm{y}},\hat{\bm{z}}$ means that components parallel (perpendicular) to the $\sharp$ direction are symmetry allowed, while the others are zero. 
All ($\bm{0}$) means that all the components are symmetry-allowed (zero).
}
\begin{ruledtabular}
\begin{tabular}{cccccccccc}
&\multicolumn{3}{c}{$J_{12}$}  & \multicolumn{3}{c}{$Y^x_{12}$} & \multicolumn{3}{c}{$Y^y_{12}$} \\
 \cline{2-4} \cline{5-7}  \cline{8-10} 
\textbf{G} & $\bm{D}_{12}$ & $\bm{E}_{12}$ & $\bm{F}_{12}$ & $\bm{A}^x_{12}$ & $\bm{B}^x_{12}$ & $\bm{C}^x_{12}$ & $\bm{A}^y_{12}$ & $\bm{B}^y_{12}$ & $\bm{C}^y_{12}$  \\ \hline
$mmm$  & $\bm{0}$ & $\bm{0}$ & All 
 & $\bm{0}$ & $\bm{0}$ & $\bm{0}$
 & $\parallel\hat{\bm{z}}$ & $\bm{0}$ & $\bm{0}$
 \\ 
$2mm$ &  $\bm{0}$ & $\bm{0}$ & All
& $\bm{0}$ & $\bm{0}$ & All
& $\parallel\hat{\bm{z}}$ & $\parallel\hat{\bm{z}}$ & $\bm{0}$ 
\\ 
$m2m$ &  $\parallel\hat{\bm{z}}$ & $\bm{0}$ & All
& $\bm{0}$ & $\parallel\hat{\bm{z}}$ & $\bm{0}$ 
& $\parallel\hat{\bm{z}}$ & $\bm{0}$ &  All
\\ 
$mm2$ &  $\parallel\hat{\bm{y}}$ & $\bm{0}$ & All
& $\bm{0}$ & $\parallel\hat{\bm{y}}$ & $\bm{0}$
&  $\parallel\hat{\bm{z}}$ & $\parallel\hat{\bm{x}}$ & $\bm{0}$
\\ 
$222$ &    $\parallel\hat{\bm{x}}$ & $\bm{0}$ & All
& $\bm{0}$ & $\parallel\hat{\bm{x}}$ & $\bm{0}$
&  $\parallel\hat{\bm{z}}$ & $\parallel\hat{\bm{y}}$ & $\bm{0}$
\\ 
$2/m..$ &  $\bm{0}$ & $\parallel\hat{\bm{x}}$ &  All
& $\parallel\hat{\bm{x}}$ & $\bm{0}$ & $\bm{0}$  
& $\perp\hat{\bm{x}}$ & $\bm{0}$ & $\bm{0}$  
\\
$.2/m.$ & $\bm{0}$ & $\parallel\hat{\bm{y}}$ & All
 & $\parallel\hat{\bm{y}}$ & $\bm{0}$ & $\bm{0}$ 
 & $\perp\hat{\bm{y}}$ & $\bm{0}$ & $\bm{0}$ 
\\ 
$..2/m$ & $\bm{0}$ & $\parallel\hat{\bm{z}}$ & All
& $\parallel\hat{\bm{z}}$ & $\bm{0}$ & $\bm{0}$ 
& $\parallel\hat{\bm{z}}$ & $\bm{0}$ & $\bm{0}$ 
\\ 
$m..$ &  $\perp\hat{\bm{x}}$ & $\parallel\hat{\bm{x}}$ &  All
& $\parallel\hat{\bm{x}}$ & $\perp\hat{\bm{x}}$ & $\bm{0}$  
& $\perp\hat{\bm{x}}$ & $\parallel\hat{\bm{x}}$ &  All
\\ 
$.m.$ & $\parallel\hat{\bm{y}}$ & $\parallel\hat{\bm{y}}$ & All
 & $\parallel\hat{\bm{y}}$ & $\parallel\hat{\bm{y}}$ & All
 & $\perp\hat{\bm{y}}$ & $\perp\hat{\bm{y}}$ & $\bm{0}$ 
 \\ 
$..m$ & $\parallel\hat{\bm{z}}$ & $\parallel\hat{\bm{z}}$ & All
& $\parallel\hat{\bm{z}}$ & $\parallel\hat{\bm{z}}$ & All
& $\parallel\hat{\bm{z}}$ & $\parallel\hat{\bm{z}}$ & All
\\ 
$2..$  & $\parallel\hat{\bm{x}}$ & $\parallel\hat{\bm{x}}$ & All
& $\parallel\hat{\bm{x}}$ & $\parallel\hat{\bm{x}}$ & All
& $\perp\hat{\bm{x}}$ & $\perp\hat{\bm{x}}$ & $\bm{0}$ 
 \\ 
$.2.$ &  $\perp\hat{\bm{y}}$ & $\parallel\hat{\bm{y}}$ & All
 &  $\parallel\hat{\bm{y}}$ & $\perp\hat{\bm{y}}$ & $\bm{0}$ 
  &  $\perp\hat{\bm{y}}$ & $\parallel\hat{\bm{y}}$ & All
 \\ 
 $..2$ & $\perp\hat{\bm{z}}$ & $\parallel\hat{\bm{z}}$ & All
 & $\parallel\hat{\bm{z}}$ & $\perp\hat{\bm{z}}$ & $\bm{0}$
&  $\parallel\hat{\bm{z}}$ & $\perp\hat{\bm{z}}$ & $\bm{0}$
\\
$\bar{1}$ &   $\bm{0}$ & All &  All
 & All & $\bm{0}$ & $\bm{0}$ 
 & All & $\bm{0}$ & $\bm{0}$ 
\end{tabular}
\end{ruledtabular}
\end{table}

\subsection{Static Hamiltonian}
\label{sec:model_static}

We consider a spin model with a generalized bilinear exchange interaction without time dependence, which is given by 
\begin{align}
\label{eq:H0}
\mathcal{H}_0=\sum_{i, j}\sum_{\alpha,\beta}J^{\alpha\beta}_{ij}S^\alpha_iS^\beta_j,
\end{align}
with
\begin{align}
\label{eq:J}
J_{ij} &= \begin{pmatrix}
F_{ij}^x & E_{ij}^z+D_{ij}^z & E_{ij}^y-D_{ij}^y \\
E_{ij}^z-D_{ij}^z & F_{ij}^y & E_{ij}^x+D_{ij}^x \\
E_{ij}^y+D_{ij}^y & E_{ij}^x-D_{ij}^x & F_{ij}^z  
\end{pmatrix}. 
\end{align}
Here, $\bm{S}_i=(S^x_i,S^y_i,S^z_i)$ is the quantum spin operator with arbitrary spin length at site $i$, the summation is taken over the
bonds in the target lattice structure, and $\alpha,\beta=x,y,z$. 
The interaction matrix $J_{ij}$ has nine independent components in each $\langle i,j \rangle$ bond: three antisymmetric off-diagonal components $\bm{D}_{ij}=(D^x_{ij},D^y_{ij},D^z_{ij})$, three symmetric off-diagonal components $\bm{E}_{ij}=(E^x_{ij},E^y_{ij},E^z_{ij})$, and three symmetric diagonal components $\bm{F}_{ij}=(F^x_{ij},F^y_{ij},F^z_{ij})$.
The antisymmetric (symmetric) components are odd (even) with respect to the interchange of two sites: $\bm{D}_{ij}=-\bm{D}_{ji}$, $\bm{E}_{ij}=\bm{E}_{ji}$, and $\bm{F}_{ij}=\bm{F}_{ji}$.
The exchange interaction with $J^{\rm iso}_{ij}=(F^x_{ij}+F^y_{ij}+F^z_{ij})/3$ is isotropic in spin space, while $D^\alpha_{ij}$, $E^\alpha_{ij}$, and $F^\alpha_{ij}-J^{\rm iso}$ are anisotropic in spin space.
The anisotropic exchange interaction originates from the relativistic spin-orbit coupling.
Among them, $\bm{D}_{ij}$ is called the DM interaction~\cite{dzyaloshinsky1958thermodynamic,moriya1960anisotropic}.

\begin{figure}[t!]
\begin{center}
\includegraphics[width=1.0\hsize]{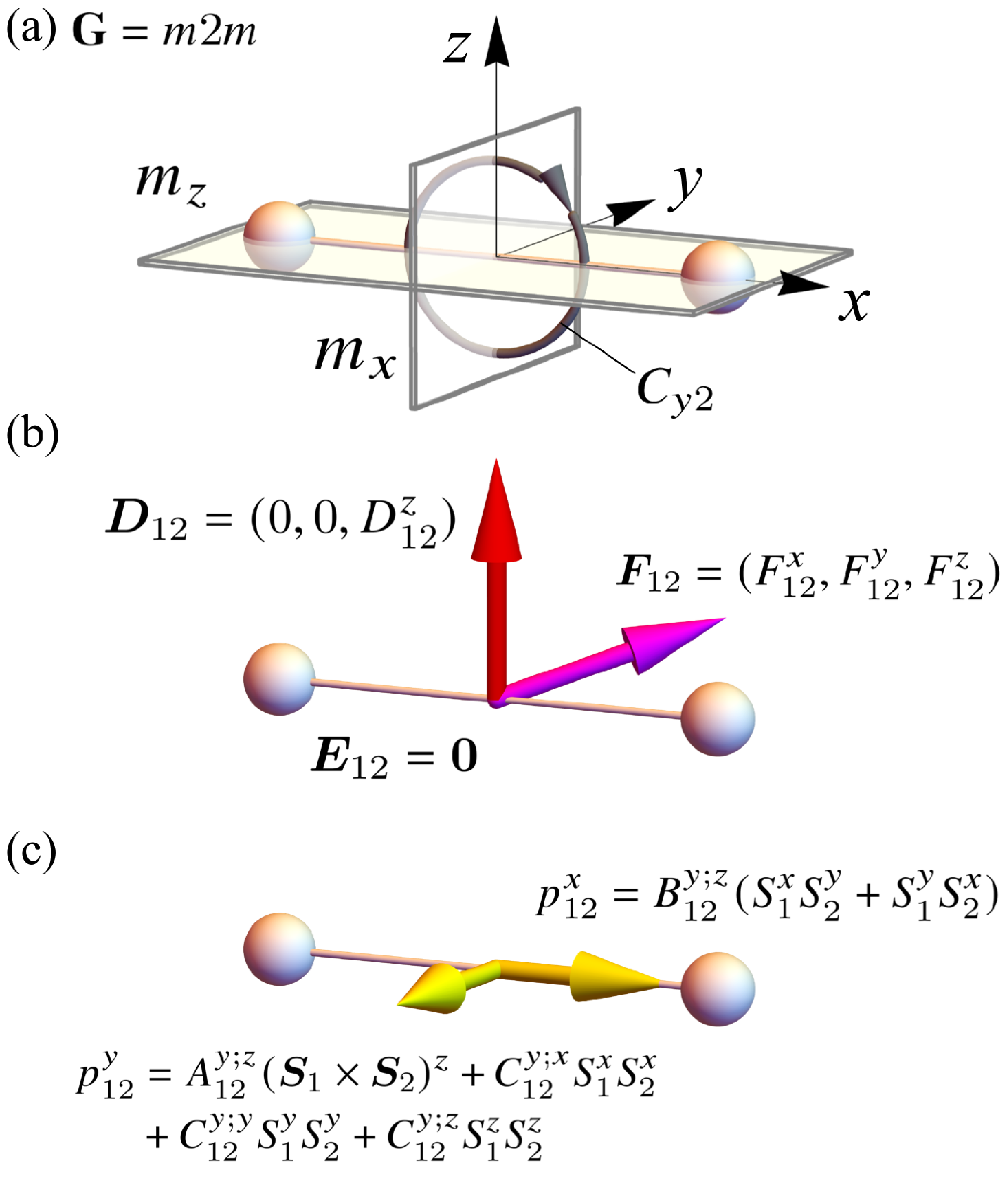} 
\caption{\label{fig:fig1}
Model parameters for (a) the $\langle 1,2 \rangle$ bond with the point group $m2m$: 
(b) $\bm{D}_{12}=(0,0,D_{12}^z)$, $\bm{E}=\bm{0}$, and $\bm{F}_{12}$ in $J_{12}$ and
(c) $\bm{A}^x_{12}=\bm{0}$, $\bm{B}^x_{12}=(0,0,B_{12}^{x;z})$, $\bm{C}^x_{12}=\bm{0}$ in $Y^x_{12}$ and $\bm{A}^y_{12}=(0,0,A_{12}^{y;z})$, $\bm{B}^y_{12}=\bm{0}$, and $\bm{C}^y_{12}$ in $Y^y_{12}$.}
\end{center}
\end{figure}

From the symmetry, the model in Eq.~(\ref{eq:H0}) is constructed once the underlying crystal structure and  symmetry of the bond are given; nonzero components in $J_{ij}$ are determined by the transformation of the $\langle i,j\rangle$ bond. 
We consider a two-dimensional system on the $xy$ plane for simplicity, where the bond symmetry on the two-dimensional plane is classified into the orthorhombic point group $mmm$ or its subgroups without two- and three-dimensional irreducible representations.  
It is noted that the following result can be extended to a three-dimensional case. 
In Table~\ref{tab:class_static}, we show the constraints on $J_{12}$ for various point groups \textbf{G}, where the $\langle 1,2 \rangle$ bond is taken along the $x$ direction.  
The symmetry of the $\langle 1,2 \rangle$ bond is classified into 15 point groups shown in Table~\ref{tab:class_static} except for the point group $\textbf{G}=1$, since point group symmetries leaving the $\langle 1,2 \rangle$ bond in the one-dimensional irreducible representation are given by a set of the space inversion ($I$), mirror perpendicular to the $\alpha=x,y,z$ axis ($m_{\alpha}$), and twofold rotation around the $\alpha$ axis ($C_{\alpha 2}$).
For example, the bond with the point group $m2m$ has $\bm{D}_{12}=(0,0,D_{12}^z)$, $\bm{E}=\bm{0}$, and $\bm{F}_{12}=(F_{12}^x,F_{12}^y,F_{12}^z)$ owing to $m_z$, $m_x$, and $C_{y2}$ symmetries, as shown in Figs.~\ref{fig:fig1}(a) and \ref{fig:fig1}(b).

Table~\ref{tab:class_static} shows that the crystal symmetry imposes constraints on $\bm{D}_{12}$ and $\bm{E}_{12}$ but not on $\bm{F}_{12}$. 
The major difference between $\bm{D}_{12}$ and $\bm{E}_{12}$ is that $\bm{D}_{12}$ becomes zero  but $\bm{E}_{12}$ is symmetry allowed under the inversion symmetry, as shown in the result for $\textbf{G}=\bar{1}$. 
The directions of $\bm{D}_{12}$ and $\bm{E}_{12}$ are determined by rotation and mirror symmetries,  as shown in the results for $\textbf{G}=2..$, $.2.$, $..2$, $m..$, $.m.$, and $..m$.
$\bm{F}_{12}$ is symmetry allowed for all the point groups, since $S^\alpha_1S^\alpha_2$ is invariant for $I$, $m_\beta$, and $C_{\beta 2}$.
Among the symmetry rules, the rules for the DM interaction is called Moriya's rule~\cite{moriya1960anisotropic}.

\subsection{Time-dependent Hamiltonian}
\label{sec:model_time}

We take into account the effect of an external circularly polarized light on the static model. 
The model in Eq.~(\ref{eq:H0}) is transformed~\cite{PhysRevLett.117.147202} as
\begin{align}
\label{eq:Ht}
\mathcal{H}(t) = \mathcal{H}_0 -\bm{E}(t)\cdot\bm{P} -\bm{B}(t)\cdot\bm{S},
\end{align}
where the time-dependent electric field $\bm{E}(t)$ and magnetic field $\bm{B}(t)$ are coupled with the electric polarization $\bm{P}$ and the total spin (magnetization) $\bm{S}=\sum_i \bm{S}_i$, respectively. 
We consider the circularly polarized light along the $z$ direction as $\bm{E}(t) = E_0(\delta\cos\Omega t,-\sin\Omega t ,0)$ and $\bm{B}(t) = B_0(-\sin\Omega,-\delta\cos\Omega t,0)$, where $\Omega$ is the light frequency and $\delta=+$ $(-)$ represents the right-circularly (left-circularly) polarized light.

In contrast to the magnetization, the expression of the electric polarization depends on the lattice geometry, since it is related to an even order of the spin product.
We consider the situation that the electric polarization originates from the spin-dependent electric dipole on the $\langle i,j \rangle$ bond ($i \neq j$)~\cite{cheong2007multiferroics,tokura2014multiferroics} as
\begin{align}
\label{eq:p}
P^\alpha &=\sum_{i, j} p^\alpha_{ij} =\sum_{i, j} \sum_{\beta,\gamma} Y^{\alpha;\beta\gamma}_{ij} S^\beta_i S^\gamma_j.
\end{align} 
Here, the polarization in the whole system is the total of the electric dipole $\bm{p}_{ij}=(p^x_{ij},p^y_{ij},p^z_{ij})$ at each $\langle i,j \rangle$ bond, where the summation with respect to $i,j$ is taken over the bonds. The $\alpha=x,y,z$ component of the electric dipole is characterized by the third-rank ME tensor $Y^{\alpha;\beta\gamma}_{ij}$, which is given by 
\begin{align} 
\label{eq:Y}
Y^\alpha_{ij} &= \begin{pmatrix}
C^{\alpha;x}_{ij} & B^{\alpha;z}_{ij}+A^{\alpha;z}_{ij} & B^{\alpha;y}_{ij}-A^{\alpha;y}_{ij} \\
B^{\alpha;z}_{ij}-A^{\alpha;z}_{ij} & C^{\alpha;y}_{ij} & B^{\alpha;x}_{ij}+A^{\alpha;x}_{ij} \\
B^{\alpha;y}_{ij}+A^{\alpha;y}_{ij} & B^{\alpha;x}_{ij}-A^{\alpha;x}_{ij} & C^{\alpha;z}_{ij}  
\end{pmatrix}.
\end{align}
The third-rank ME tensor consists of antisymmetric off-diagonal components $\bm{A}^\alpha_{ij}=(A^{\alpha;x}_{ij},A^{\alpha;y}_{ij},A^{\alpha;z}_{ij})$, symmetric off-diagonal components $\bm{B}^\alpha_{ij}=(B^{\alpha;x}_{ij},B^{\alpha;y}_{ij},B^{\alpha;z}_{ij})$, and symmetric diagonal components $\bm{C}^\alpha_{ij}=(C^{\alpha;x}_{ij},C^{\alpha;y}_{ij},C^{\alpha;z}_{ij})$, where they satisfy $\bm{A}^\alpha_{ij}=-\bm{A}^\alpha_{ji}$, $\bm{B}^\alpha_{ij}=\bm{B}^\alpha_{ji}$, and $\bm{C}^\alpha_{ij}=\bm{C}^\alpha_{ji}$ with respect to the interchange of two sites.
The polarization mechanism for $\bm{A}^\alpha_{ij}$ is an extension of the inverse DM (spin current) mechanism~\cite{Katsura_PhysRevLett.95.057205,Mostovoy_PhysRevLett.96.067601,SergienkoPhysRevB.73.094434,Harris_PhysRevB.73.184433,PhysRevB.76.054447}, $\bm{p}_{ij}\propto\bm{e}_{ij}\times (\bm{S}_i\times\bm{S}_j)$ with the bond vector $\bm{e}_{ij}$. 
Meanwhile, the mechanism based on the symmetric components $\bm{C}^\alpha_{ij}$ includes the exchange striction mechanism described by $C^{\alpha;x}_{ij}=C^{\alpha;y}_{ij}=C^{\alpha;z}_{ij}$~\cite{hur2004electric,choi2008ferroelectricity}.
Similarly to the coupling matrix $J_{ij}$, zero components in $Y^\alpha_{ij}$ are determined by the symmetry of the bond, as summarized for $Y^x_{12}$ and $Y^y_{12}$ under 15 point groups in Table~\ref{tab:class_static}~\cite{PhysRevB.83.174432, matsumoto2017symmetry}; it is noted that there is no $z$-directional polarization, i.e., $Y^z_{12}=0$ owing to $E^z(t)=0$. 
In Fig.~\ref{fig:fig1}(c), we show the spin-dependent electric dipoles under the $\textbf{G}=m2m$ symmetry as an example.

The symmetry rules for $Y^x_{12}$ and $Y^y_{12}$ are different from those for $J_{12}$ owing to the different symmetry in the left-hand side in Eqs.~(\ref{eq:H0}) and (\ref{eq:p}); 
$\mathcal{H}_0$ corresponds to the scalar belonging to the totally symmetric irreducible representation and $P^{\alpha}$ corresponds to the polar vector belonging to the different irreducible representation from $\mathcal{H}_0$.
As a result, the inversion symmetry forbids (allow) the antisymmetric components $\bm{D}_{12}$ (symmetric ones $\bm{E}_{12}$ and $\bm{F}_{12}$) in $J_{12}$ and the symmetric ones $\bm{B}^\alpha_{12}$ and $\bm{C}^\alpha_{12}$ (antisymmetric one $\bm{A}^\alpha_{12}$) in $Y^x_{12}$ and $Y^y_{12}$; see $\textbf{G}=\bar{1}$ as an example.
The directions of $\bm{A}^\alpha_{12}$, $\bm{B}^\alpha_{12}$, and $\bm{C}^\alpha_{12}$ are determined by rotation and mirror symmetries, where $A^{y;z}_{12}$ exists for all the point groups, since $p^y_{12}$ and $A^{y;z}_{12}(S^x_1S^y_2-S^y_2S^x_1)$ have the same symmetry. 
In other words, $A^{y;z}_{12}$ in the third rank ME tensor is the counterpart of $\bm{F}_{12}$ in the interaction matrix.

\subsection{Effective time-independent Hamiltonian}
\label{sec:model_effective}

We show specific expressions of  light-induced anisotropic exchange interactions in Eq.~(\ref{eq:Ht}) based on the Floquet theory.
 \cite{eckardt2015high,PhysRevB.93.144307}. 
The effective Hamiltonian up to the first order of $\Omega^{-1}$ is given by
\begin{align} 
\label{eq: Hameff1}
\mathcal{H}_\mathrm{eff} &= \mathcal{H}_0 + \frac{1}{\Omega}\sum_{m>0}\frac{[H_{-m},H_{+m}]}{m}.
\end{align} 
Here, $H_m$ is the Fourier transform of the time-periodic Hamiltonian, $H(t)=\sum_m e^{-im\Omega t}H_m$ with integer $m$, and $[H_{-m},H_{+m}]$ is the commutation relation.
Since the light-induced modulation in the second term does not depend on $\mathcal{H}_0$ within the order of $\Omega^{-1}$, the following results are not affected by the original static Hamiltonian.

By substituting Eq.~(\ref{eq:Ht}) into Eq.~(\ref{eq: Hameff1}), the effective Hamiltonian is given by~\cite{PhysRevLett.117.147202}
\begin{align} 
\label{eq:Heff}
\mathcal{H}_\mathrm{eff} &= \mathcal{H}_0 +\mathcal{H}_\mathrm{1 spin} + \mathcal{H}_\mathrm{2 spin} +\mathcal{H}_\mathrm{3 spin},
\end{align} 
with
\begin{align}
\label{eq:H1}
\mathcal{H}_\mathrm{1spin} &= -\frac{i\delta B_0^2 }{2\Omega}[S^x,S^y]= \frac{\delta B_0^2 }{2\Omega} S^z, \\
\label{eq:H2}
\mathcal{H}_\mathrm{2spin} &= -\frac{i\delta E_0B_0}{2\Omega} ( [P^x,S^x] + [P^y,S^y] ),\\
\label{eq:H3}
\mathcal{H}_\mathrm{3spin} &= -\frac{i\delta E_0^2}{2\Omega} [P^x,P^y].
\end{align}
$\mathcal{H}_\mathrm{1spin}$ corresponds to an effective magnetic field along the $z$ direction coupled with $S^z$ and is irrespective of the third-rank ME tensor; there is no dependence on the crystal structure. 
Meanwhile, $\mathcal{H}_\mathrm{2spin}$ and $\mathcal{H}_\mathrm{3spin}$ depend on the third-rank ME tensor, which include the information about the crystal structure.
In other words, $\mathcal{H}_\mathrm{2spin}$ and $\mathcal{H}_\mathrm{3spin}$ give rise to a variety of light-induced magnetic interactions depending on the crystal symmetry. 
Indeed, previous studies have shown that the DM interaction is induced in $\mathcal{H}_\mathrm{2spin}$ via the inverse DM mechanism~\cite{PhysRevLett.117.147202}, a three-spin interaction related to the spin scalar chirality is induced in $\mathcal{H}_\mathrm{3spin}$ via the exchange striction mechanism on a honeycomb lattice~\cite{sato2014floquet}, and a single-ion anisotropy is induced in $\mathcal{H}_\mathrm{3spin}$ via the $d$-$p$ hybridization mechanism~\cite{PhysRevLett.128.037201}.  
Hereafter, we present a systematic classification of the light-induced magnetic interactions based on the general third-rank ME tensor that is applicable to any two-dimensional systems.

Let us comment on details of the light-induced magnetic interactions~\cite{PhysRevLett.117.147202}. 
First, the magnetic interactions in Eqs.~(\ref{eq:H1})-(\ref{eq:H3}) can be induced even though the light is elliptically polarized, while their amplitudes become smaller than those by the circularly 
polarized light.
Second, the light-induced magnetic interactions work for a finite time (Floquet prethermal regime), and then they stop modulating magnetic structures.
Finally, although the magnetic interactions in Eqs.~(\ref{eq:H1})-(\ref{eq:H3}) are obtained based on the quantum nature of the spin, they can be used to investigate the classical spin dynamics by solving the Landau-Lifshitz-Gilbert (LLG) equation for $\mathcal{H}_\mathrm{eff} $ in Eq.~(\ref{eq:Heff}) with the classical spin~\cite{PhysRevLett.128.037201,higashikawa2018floquet}.

\section{Symmetry analysis via Floquet theory}
\label{sec:symmetry}

\begin{table}
\caption{
\label{tab:group}
Change of point group \textbf{G}.
$\mathcal{H}_\mathrm{2 spin}$ ($\mathcal{H}_\mathrm{3 spin}$) changes \textbf{G} into the chiral point group \textbf{G}$^\mathrm{(C)}$ (black and white magnetic point group \textbf{M}).
$\mathcal{H}_\mathrm{2 spin}$ and $\mathcal{H}_\mathrm{3 spin}$ change \textbf{G} into chiral black and white magnetic point groups \textbf{M}$^\mathrm{(C)}$.
The 1st, 2nd, and 3rd axes are [100], [010], and [001], respectively.
}
\begin{ruledtabular}
\begin{tabular}{cccc}
\textbf{G} & \textbf{G}$^\mathrm{(C)}$ & \textbf{M} & \textbf{M}$^\mathrm{(C)}$ \\
\hline
$mmm$ & $222$ & $m'm'm$ & $2'2'2$
\\ 
$2mm$ & $2..$ & $2'm'm$ & $2'..$
\\
$m2m$ & $.2.$ & $m'2'm$ & $.2'.$
 \\
 $mm2$ & $..2$ & $m'm'2$ & $..2$
\\
$222$ & $222$ & $2'2'2$ & $2'2'2$
\\
$2/m..$ & $2..$ & $2'/m'..$ & $2'..$
\\ 
$.2/m.$ & $.2.$ & $.2'/m'.$ & $.2'.$
\\
$..2/m$ & $..2$ & $..2/m$ & $..2$
\\
$m..$ & $1$ & $m'..$ & $1$
\\
$.m.$ & $1$ & $.m'.$ & $1$
\\
$..m$ & $1$ & $..m$ & $1$
\\
$2..$ & $2..$ & $2'..$ & $2'..$
 \\
$.2.$ & $.2.$ & $.2'.$ & $.2'.$
\\
$..2$ & $..2$ & $..2$ & $..2$
\\
$\bar{1}$ & $1$ &$\bar{1}$ & $1$
\end{tabular}
\end{ruledtabular}
\end{table}

We systematically derive a variety of light-indued magnetic interactions depending on crystal structures: anisotropic two-site two-spin interactions in Sec.~\ref{sec:2spin}, anisotropic two-site three-spin interactions in Sec.~\ref{sec:Q}, and anisotropic three-site three-spin interactions in Sec.~\ref{sec:3site}. 
The former two-spin interaction arises from $\mathcal{H}_\mathrm{2spin}$, while the latter three-spin interactions arise from $\mathcal{H}_\mathrm{3spin}$.
Our symmetry analysis clarifies that the appearance of the anisotropic two-site two-spin interactions (anisotropic two-site three-spin and three-site three-spin interactions) is owing to the reduction of the point group \textbf{G} to chiral point group \textbf{G}$^\mathrm{(C)}$ (black and white magnetic point group \textbf{M}). 
We show such a correspondence in Table~\ref{tab:group}.

\subsection{Anisotropic two-site two-spin interaction}
\label{sec:2spin}   

First, we show the general expression of the light-induced two-site two-spin interaction that originates from the commutation of the electric polarization $\bm{P}$ and the magnetization $\bm{S}$ in Eq.~(\ref{eq:H2}). 
By substituting the third-rank ME tensor in Eq.~(\ref{eq:Y}) into Eq.~(\ref{eq:H2}), we obtain the light-induced two-site two-spin interaction given by  
\begin{align}
\mathcal{H}_\mathrm{2 spin} &= \sum_{i, j} \sum_{\alpha,\beta} \mathcal{J}^{\alpha\beta}_{ij}S^\alpha_i S^\beta_j,
\end{align}
with
\begin{align}
\mathcal{J}_{ij} &= \begin{pmatrix}
\mathcal{F}_{ij}^x & \mathcal{E}_{ij}^z+\mathcal{D}_{ij}^z & \mathcal{E}_{ij}^y-\mathcal{D}_{ij}^y \\
\mathcal{E}_{ij}^z-\mathcal{D}_{ij}^z & \mathcal{F}_{ij}^y & \mathcal{E}_{ij}^x+\mathcal{D}_{ij}^x \\
\mathcal{E}_{ij}^y+\mathcal{D}_{ij}^y & \mathcal{E}_{ij}^x-\mathcal{D}_{ij}^x & \mathcal{F}_{ij}^z 
\end{pmatrix}.
\end{align}
Here, the summation is taken over the bonds; $\bm{\mathcal{D}}_{ij}=(\mathcal{D}^x_{ij},\mathcal{D}^y_{ij},\mathcal{D}^z_{ij})$ are light-induced DM interactions, $\bm{\mathcal{E}}_{ij}=(\mathcal{E}^x_{ij},\mathcal{E}^y_{ij},\mathcal{E}^z_{ij})$ are light-induced symmetric off-diagonal interactions, and $\bm{\mathcal{F}}_{ij}=(\mathcal{F}^x_{ij},\mathcal{F}^y_{ij},\mathcal{F}^z_{ij})$ are light-induced symmetric diagonal interactions.
The coupling matrix $\mathcal{J}_{ij}$ is related to the third-rank ME tensor as
\begin{align}
\label{eq:Dx}
\frac{2\Omega}{\delta E_0B_0}\mathcal{D}^x_{ij} & =  -A^{y;z}_{ij}, \\ 
\frac{2\Omega}{\delta E_0B_0}\mathcal{D}^y_{ij} & =  A^{x;z}_{ij}, \\ 
\label{eq:Dz}
\frac{2\Omega}{\delta E_0B_0}\mathcal{D}^z_{ij} & =-A^{x;y}_{ij} + A^{y;x}_{ij}, \\
\frac{2\Omega}{\delta E_0B_0}\mathcal{E}^x_{ij} & = -C^{x;y}_{ij} + C^{x;z}_{ij} + B^{y;z}_{ij}, \\
\frac{2\Omega}{\delta E_0B_0}\mathcal{E}^y_{ij} & = -C^{y;z}_{ij} + C^{y;x}_{ij} - B^{x;z}_{ij}, \\
\frac{2\Omega}{\delta E_0B_0}\mathcal{E}^z_{ij} & =B^{x;y}_{ij} - B^{y;x}_{ij}, \\ 
\frac{2\Omega}{\delta E_0B_0}\mathcal{F}^x_{ij} & = -2B^{y;y}_{ij}, \\ 
\frac{2\Omega}{\delta E_0B_0}\mathcal{F}^y_{ij} & = 2B^{x;x}_{ij}, \\ 
\label{eq:Fz}
\frac{2\Omega}{\delta E_0B_0}\mathcal{F}^z_{ij} & = -2B^{x;x}_{ij} + 2B^{y;y}_{ij},
\end{align}
where the antisymmetric (symmetric) interactions are induced by the antisymmetric (symmetric) components in the third-rank ME tensor since the parity with respect to the interchange of two sites in $\mathcal{H}_\mathrm{2 spin}$ linearly depends on $\bm{p}_{ij}$.   
Equations~(\ref{eq:Dx})-(\ref{eq:Fz}) show that all the types of anisotropic two-site two-spin interactions can be induced by the light depending on the symmetry of the $\langle i,j \rangle$ 
bond.

\begin{table}
\caption{
\label{tab:class_H2}
Nonzero components in $\mathcal{J}_{12}$ under point group \textbf{G}.
Components with $\checkmark$ mean the light-induced ones via $Y^x_{12}$ and $Y^y_{12}$ shown in Table~\ref{tab:class_static}.
}
\begin{ruledtabular}
\begin{tabular}{cccccccccc}
\textbf{G} & $\mathcal{D}^x_{12}$ & $\mathcal{D}^y_{12}$ & $\mathcal{D}^z_{12}$ & $\mathcal{E}^x_{12}$ & $\mathcal{E}^y_{12}$ & $\mathcal{E}^z_{12}$ & $\mathcal{F}^x_{12}$ & $\mathcal{F}^y_{12}$ & $\mathcal{F}^z_{12}$  \\ \hline
$mmm$ &$\checkmark$ & & & & & & & &
\\ 
$2mm$ &$\checkmark$ & & &$\checkmark$ & & & & & 
\\
$m2m$ &$\checkmark$ & & & &$\checkmark$ & & & & 
 \\
$mm2$ &$\checkmark$ & & & & &$\checkmark$ & & &
\\
$222$ &$\checkmark$ & & & & & &$\checkmark$ &$\checkmark$ &$\checkmark$ 
\\
$2/m..$ &$\checkmark$ & & & & & & & & 
\\ 
$.2/m.$ &$\checkmark$ & &$\checkmark$ & & & & & & 
\\
$..2/m$ &$\checkmark$ &$\checkmark$ & & & & & & &   
\\
$m..$ &$\checkmark$ & & & &$\checkmark$ &$\checkmark$ & & & 
\\
$.m.$ &$\checkmark$ & &$\checkmark$ &$\checkmark$ & &$\checkmark$ & & &
\\
$..m$ &$\checkmark$ &$\checkmark$ & &$\checkmark$ &$\checkmark$ & & & & 
\\
$2..$ &$\checkmark$ & & &$\checkmark$ & & &$\checkmark$ &$\checkmark$ &$\checkmark$ 
\\
$.2.$ &$\checkmark$ & &$\checkmark$ & &$\checkmark$ & &$\checkmark$ &$\checkmark$ &$\checkmark$  
\\
$..2$ &$\checkmark$ &$\checkmark$ & & & &$\checkmark$ &$\checkmark$ &$\checkmark$ &$\checkmark$   
\\
$\bar{1}$ &$\checkmark$ &$\checkmark$ &$\checkmark$ & 
\end{tabular}
\end{ruledtabular}
\end{table}

To investigate the crystal-structure dependence of the light-induced anisotropic two-site two-spin interaction, we calculate nonzero components in $\mathcal{J}_{12}$ for each point group, as summarized in Table~\ref{tab:class_H2}.
These results are obtained by substituting the symmetry-allowed ME components shown in Table~\ref{tab:class_static} into Eqs.~(\ref{eq:Dx})-(\ref{eq:Fz}). 
We find two major characteristics.
One is the absence of the light-induced symmetric interactions under point groups with the inversion symmetry, $mmm$, $2/m..$, $.2/m.$, $..2/m$, and $\bar{1}$, since the symmetric components in the third-rank ME tensor vanish under the inversion symmetry.
The other is the presence of the light-induced chiral-type DM interaction $\mathcal{D}^x_{12}$ irrespective of the point group, since $A^{y;z}_{12}$ is symmetry allowed for all the point groups.
In other words, the light-induced DM interactions are present irrespective of  the inversion symmetry, where they originate from $\bm{A}^{x}_{12}$ and $\bm{A}^{y}_{12}$ unrelated to the inversion symmetry, as shown in Eqs.~(\ref{eq:Dx})-(\ref{eq:Dz}) and Table~\ref{tab:class_static}.
From these observations, one can find the opposite tendency of the symmetry-allowed interactions between the light-induced two-site two-spin interaction $\mathcal{J}_{ij}$
and that of the static one $J_{ij}$ in Eq.~(\ref{eq:H0}); the inversion symmetry forbids (allows) the static (light-induced) DM interaction, while the inversion symmetry allows (forbids) the static (light-induced) symmetric interaction. 
It means that magnetic structures favored by the static DM interaction in noncentrosymmetric systems, such as a spiral state and skyrmion crystal, can be realized in centrosymmetric systems by applying the circularly polarized light.    

\begin{figure}[t!]
\begin{center}
\includegraphics[width=1.0\hsize]{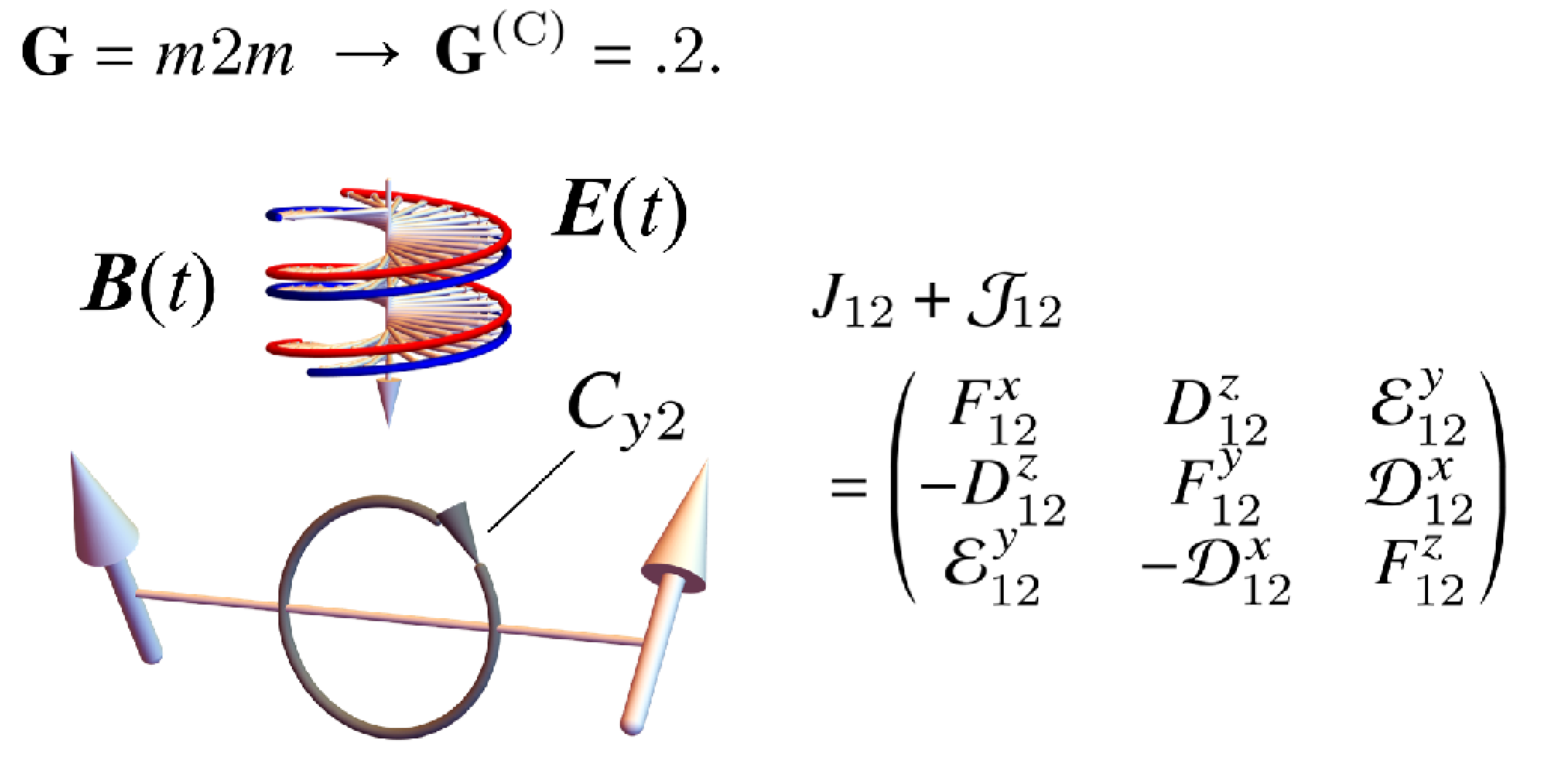} 
\caption{\label{fig:fig2}
The static and light-induced two-site two-spin interactions, where the point group $\textbf{G}=m2m$ shown in Fig.~\ref{fig:fig1}(a) is lowered to the chiral point group $\textbf{G}^{\rm (C)}=.2.$ by the electric field $\bm{E}(t)$ (red) and the magnetic field $\bm{B}(t)$ (blue).
}
\end{center}
\end{figure}

We find that the appearance of the light-induced two-site two-spin interactions corresponds to the change of the point group \textbf{G} under $\mathcal{H}_\mathrm{2 spin}\propto i( [P^x,S^x] + [P^y,S^y] )$.
Since $\bm{P}$ is space-inversion odd and time-reversal even, while $i\bm{S}$ is space-inversion even and time-reversal even, $\mathcal{H}_\mathrm{2 spin}$ holds the time-reversal symmetry but breaks the inversion symmetry $I$.
Furthermore, the opposite parity for the mirror symmetry of $\bm{P}$ and $\bm{S}$ means the breaking of all the mirror symmetries $m_x$, $m_y$, and $m_z$ with keeping all the rotational symmetry $C_{x2}$, $C_{y2}$, and $C_{z2}$. 
Accordingly, the point group \textbf{G} changes into the chiral point group \textbf{G}$^\mathrm{(C)}$ by $\mathcal{H}_\mathrm{2 spin}$ to accommodate the coupling between the polar electric and axial magnetic fields~\cite{kishine2022definition}.
We show such a symmetry lowering in Table.~\ref{tab:group}, where \textbf{G}$^\mathrm{(C)}$ is obtained by extracting the symmetries $\{I,m_x,m_y,m_z\}$ from \textbf{G}. 
We confirm that the modulation of the anisotropic two-spin interaction by light based on the perturbation argument is consistent with the reduction to the chiral point groups: When considering the total coupling matrix $J^\mathrm{total}_{12}=J_{12}+\mathcal{J}_{12}$, zero components in $J^\mathrm{total}_{12}$ are determined by the chiral point group $\textbf{G}^\mathrm{(C)}$\footnote{When the point groups do not have symmetries $\{I,m_x,m_y,m_z\}$ in \textbf{G}, such as $222$, $2..$, $.2.$, and $..2$, no symmetry reduction occurs; no additional two-spin interaction appears but the static two-spin interactions are modified.}.
.
For example, the point group $m2m$ [Fig.~\ref{fig:fig1}(a)] is reduced to the chiral point group $.2.$ (Fig.~\ref{fig:fig2}).  
Then, $J^\mathrm{total}_{12}$ has light-induced $\mathcal{D}^x_{12}$ and $\mathcal{E}^{y}_{12}$ in addition to static $\bm{F}_{12}$ and $D^z_{12}$, all of which are allowed by $\textbf{G}^\mathrm{(C)}=.2.$ (see Table.~\ref{tab:class_static}).

\subsection{Anisotropic two-site three-spin interaction}
\label{sec:Q}   

The anisotropic two-site three-spin interaction is obtained from $\mathcal{H}_\mathrm{3 spin}$ as 
\begin{align} 
\mathcal{H}_\mathrm{Q}&= -\frac{i\delta E_0^2}{2\Omega}\sum_{i,j} [p^x_{ij},p^y_{ij}] \nonumber\\
\label{eq:HQ}
 &= 
\sum_{i,j} \sum_{\alpha,\beta,\gamma}\left\{ \mathcal{O}^{(\mathrm{S})\alpha\beta\gamma}_{ij}(Q^{\alpha\beta}_iS^\gamma_j +S^\gamma_i Q^{\alpha\beta}_j) \right. \nonumber\\
 &\left.+\mathcal{O}^{(\mathrm{AS})\alpha\beta\gamma}_{ij}(Q^{\alpha\beta}_iS^\gamma_j -S^\gamma_i Q^{\alpha\beta}_j)  \right\}, 
\end{align}
where the summation is taken over the bonds and we introduce electric quadrupoles $Q^{\alpha\beta}_i$ at site $i$, which is defined as 
\begin{align}
\label{eq:quadrupole}
Q^{\beta\gamma}_i=\frac{1}{2}(S^\beta_iS^\gamma_i+S^\gamma_iS^\beta_i).
\end{align}
Thus, the two-site three-spin interaction is regarded as a spin-quadrupole interaction between two sites.
As well as the anisotropic two-site two-spin interaction, the anisotropic two-site three-spin interaction is divided into even and odd components with respect to the interchange of two sites: symmetric ones satisfying $\mathcal{O}^{(\mathrm{S})\alpha\beta\gamma}_{ij}=\mathcal{O}^{(\mathrm{S})\alpha\beta\gamma}_{ji}$ and antisymmetric ones satisfying $\mathcal{O}^{(\mathrm{AS})\alpha\beta\gamma}_{ij}=-\mathcal{O}^{(\mathrm{AS})\alpha\beta\gamma}_{ji}$.
In addition, $\mathcal{O}^{(\mathrm{S})\alpha\beta\gamma}_{ij}=\mathcal{O}^{(\mathrm{S})\beta\alpha\gamma}_{ij}$ and $\mathcal{O}^{(\mathrm{AS})\alpha\beta\gamma}_{ij}=\mathcal{O}^{(\mathrm{AS})\beta\alpha\gamma}_{ij}$ hold due to the quadrupole nature.  

To derive a general expression of the two-site three-spin interactions in terms of the third-rank ME tensor,
we rewrite vectors of the ME couplings as $\tilde{\bm{A}}^\alpha_{ij}=(A^{x;\alpha}_{ij},A^{y;\alpha}_{ij},0)$, $\tilde{\bm{B}}^\alpha_{ij}=(B^{x;\alpha}_{ij},B^{y;\alpha}_{ij},0)$, $\tilde{\bm{C}}^\alpha_{ij}=(C^{x;\alpha}_{ij},C^{y;\alpha}_{ij},0)$ from Eq.~(\ref{eq:Y}), where $\tilde{\bm{A}}^\alpha_{ij}$ ($\tilde{\bm{B}}^\alpha_{ij}$ and $\tilde{\bm{C}}^\alpha_{ij}$) is antisymmetric (symmetric) with respect to the interchange of two sites.  
Then, the symmetric two-site three-spin interactions are given by
\begin{align}
\frac{2\Omega}{\delta E_0^2}\mathcal{O}^{(\mathrm{S})\alpha\alpha\alpha}_{ij} &= \frac{1}{2}\sum_{\beta,\gamma}\epsilon_{\alpha\beta\gamma}(\tilde{\bm{A}}^{\beta}_{ij} \times\tilde{\bm{A}}^{\gamma}_{ij}-\tilde{\bm{B}}^{\beta}_{ij} \times\tilde{\bm{B}}^{\gamma}_{ij}  )^z, \\
\frac{2\Omega}{\delta E_0^2}\mathcal{O}^{(\mathrm{S})\alpha\alpha\beta}_{ij} &= -\sum_{\gamma}\epsilon_{\alpha\beta\gamma}(\tilde{\bm{C}}^{\alpha}_{ij} \times\tilde{\bm{B}}^{\beta}_{ij} )^z, \\
\frac{2\Omega}{\delta E_0^2}\mathcal{O}^{(\mathrm{S})\alpha\beta\gamma}_{ij} &= \epsilon_{\alpha\beta\gamma}(\tilde{\bm{C}}^{\alpha}_{ij} \times\tilde{\bm{C}}^{\beta}_{ij} )^z \\
&+ \delta_{\alpha\gamma}\sum_{\eta}\epsilon_{\gamma\beta\eta}(\tilde{\bm{B}}^{\eta}_{ij} \times\tilde{\bm{B}}^{\gamma}_{ij} \nonumber\\
&+\tilde{\bm{A}}^{\eta}_{ij} \times\tilde{\bm{A}}^{\gamma}_{ij}+\tilde{\bm{C}}^{\beta}_{ij} \times\tilde{\bm{B}}^{\beta}_{ij} )^z \nonumber\\
&+ \delta_{\beta\gamma}\sum_{\eta}\epsilon_{\gamma\alpha\eta}(\tilde{\bm{B}}^{\eta}_{ij} \times\tilde{\bm{B}}^{\gamma}_{ij} \nonumber\\
&+\tilde{\bm{A}}^{\eta}_{ij} \times\tilde{\bm{A}}^{\gamma}_{ij}+\tilde{\bm{C}}^{\alpha}_{ij} \times\tilde{\bm{B}}^{\alpha}_{ij} )^z,
\end{align}
where $\epsilon_{\alpha\beta\gamma}$ is the Levi-Civita symbol and the $z$ components of the outer product of the vectors originate from $ [p^x_{ij},p^y_{ij}]$.
There is no contribution from $\tilde{\bm{A}}^{\alpha}_{ij} \times\tilde{\bm{B}}^{\beta}_{ij}$ and  $\tilde{\bm{A}}^{\alpha}_{ij} \times\tilde{\bm{C}}^{\beta}_{ij}$, since such products are antisymmetric with respect to interchange of two sites.
By substituting the symmetry-allowed ME components shown in Table~\ref{tab:class_static} into the general expression, we calculate nonzero components of the symmetric two-site three-spin interactions induced in each point group, as summarized in Tables~\ref{tab:class_OS}.  

 The product of symmetric and antisymmetric ME components gives rise to the antisymmetric two-site three-spin interactions as
\begin{align}
\frac{2\Omega}{\delta E_0^2}\mathcal{O}^{(\mathrm{AS})\alpha\alpha\alpha}_{ij} &= \frac{1}{2}\sum_{\beta,\gamma}|\epsilon_{\alpha\beta\gamma}|( \tilde{\bm{A}}^{\beta}_{ij} \times\tilde{\bm{B}}^{\gamma}_{ij}+\tilde{\bm{A}}^{\gamma}_{ij} \times\tilde{\bm{B}}^{\beta}_{ij} )^z, \\
\frac{2\Omega}{\delta E_0^2}\mathcal{O}^{(\mathrm{AS})\alpha\alpha\beta}_{ij} &=(1-\delta_{\alpha\beta})(\tilde{\bm{C}}^{\alpha}_{ij} \times\tilde{\bm{A}}^{\beta}_{ij} )^z, \\
\frac{2\Omega}{\delta E_0^2}\mathcal{O}^{(\mathrm{AS})\alpha\beta\gamma}_{ij} &=  2|\epsilon_{\alpha\beta\gamma}|(\tilde{\bm{B}}^{\gamma}_{ij} \times\tilde{\bm{A}}^{\gamma}_{ij} )^z \nonumber\\
&+\delta_{\alpha\gamma}\sum_{\eta}|\epsilon_{\gamma\beta\eta}|(\tilde{\bm{B}}^{\eta}_{ij} \times\tilde{\bm{A}}^{\gamma}_{ij}\nonumber\\
&+\tilde{\bm{A}}^{\eta}_{ij} \times\tilde{\bm{B}}^{\gamma}_{ij} -\tilde{\bm{C}}^{\beta}_{ij} \times\tilde{\bm{A}}^{\beta}_{ij} )^z \nonumber\\
&+\delta_{\beta\gamma}\sum_{\eta}|\epsilon_{\gamma\alpha\eta}|(\tilde{\bm{B}}^{\eta}_{ij} \times\tilde{\bm{A}}^{\gamma}_{ij}\nonumber\\
&+\tilde{\bm{A}}^{\eta}_{ij} \times\tilde{\bm{B}}^{\gamma}_{ij} -\tilde{\bm{C}}^{\alpha}_{ij} \times\tilde{\bm{A}}^{\alpha}_{ij} )^z.
\end{align}
These general expressions mean that the antisymmetric two-site three-spin interactions vanish on the inversion-symmetric bond, where all the symmetric ME components are zero, as discussed in Sec.~\ref{sec:model_static}.    
The antisymmetric two-site three-spin interactions induced in each point group are shown in Tables~\ref{tab:class_OAS}.  

\begin{figure}[t!]
\begin{center}
\includegraphics[width=1.0\hsize]{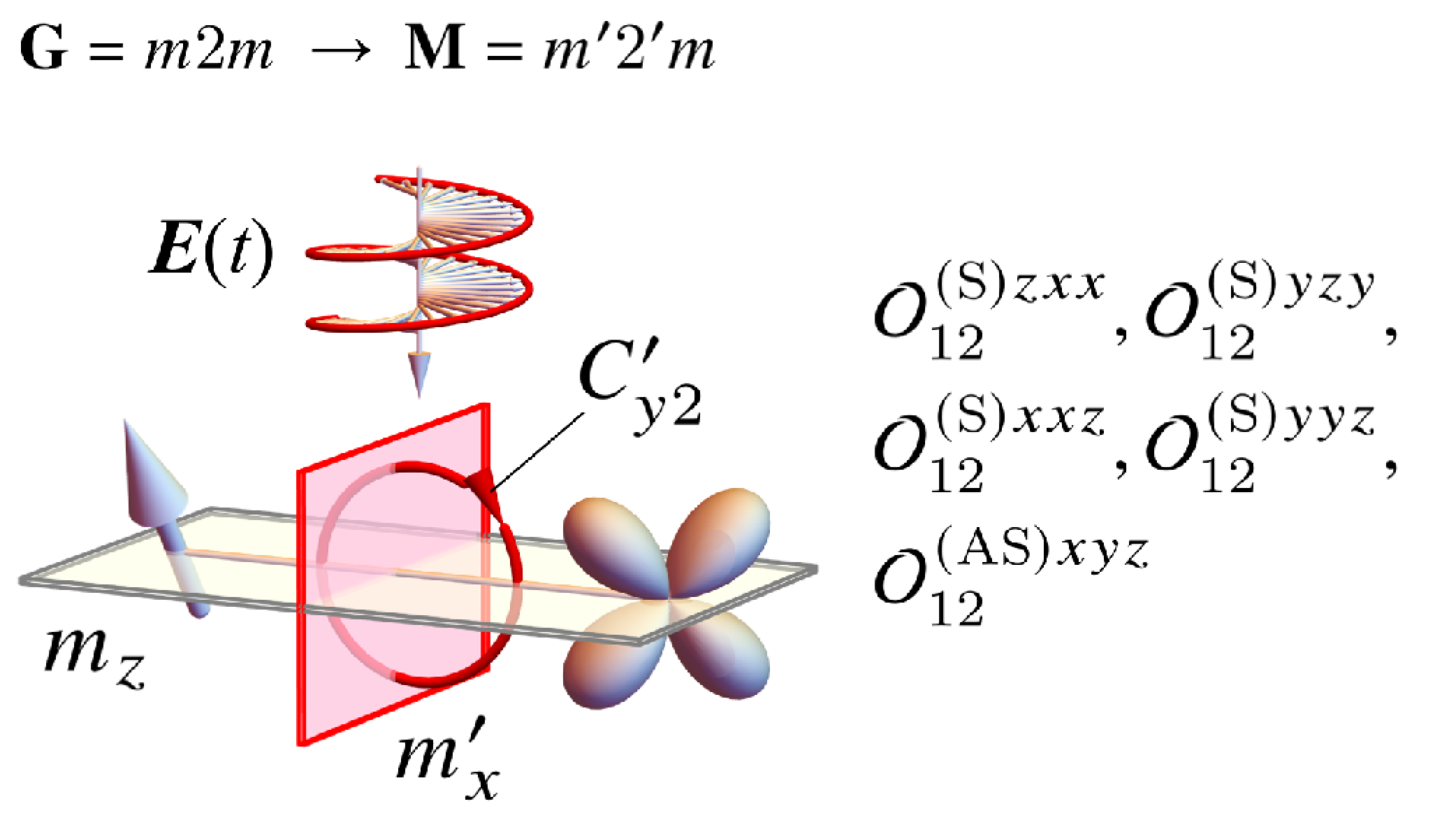} 
\caption{\label{fig:fig3}
The light-induced two-site three-spin interactions, where the reduction from the point group $\textbf{G}=m2m$ shown in Fig.~\ref{fig:fig1}(a) to the black and white magnetic point group $\textbf{M}=m'2'm$ occurs by the electric field $\bm{E}(t)$}.
\end{center}
\end{figure}

To understand the light-induced two-site three-spin interactions based on the crystal symmetry, we investigate symmetry reduction from the point group \textbf{G} under $\mathcal{H}_\mathrm{3 spin}$.
As shown in Eq.~(\ref{eq:H3}), $\mathcal{H}_\mathrm{3 spin}$ is proportional to $i[P^x,
P^y]$, which has the same symmetry as the magnetization along the $z$ direction.
The $z$ magnetization breaks the point group symmetries $\{m_x,m_y,C_{2x},C_{2y}\}$ but holds these point group symmetries combined with the time-reversal operation denoted as $\{m'_x,m'_y,C'_{2x},C'_{2y}\}$.
Thus, $\mathcal{H}_\mathrm{3 spin}$ changes the point group \textbf{G} into the black and white magnetic point group \textbf{M}~\cite{bradley2009mathematical}, as shown in Table~\ref{tab:group}; \textbf{M} and \textbf{G} in Table~\ref{tab:group} are related as $\textbf{M}=(\textbf{G}-\bar{\rm G})+\theta \bar{\rm G}$, where $\bar{\rm G}=\textbf{G}\cap\{m_x,m_y,C_{2x},C_{2y}\}$ and $\theta$ is the time-reversal operation\footnote{When the point groups do not have symmetries $\{m_x,m_y,C_{2x},C_{2y}\}$ in \textbf{G}, such as $..2/m$, $..m$, $..2$, and $\bar{1}$, no symmetry reduction to the black and white magnetic point group occurs; the three-spin interactions appear through the breaking of the time-reversal symmetry.}.
We show a symmetry rule for the two-site three-spin interactions in Table~\ref{tab:class_M} and confirm that the emergent interactions in Tables~\ref{tab:class_OS} and \ref{tab:class_OAS} satisfy the symmetry rule.
For example, the point group $m2m$ [Fig.~\ref{fig:fig1}(a)] is reduced to the black and white magnetic point group $m'2'm$ (Fig.~\ref{fig:fig3}), and $\mathcal{O}_{12}^{{\rm (S)}zxx}$,  $\mathcal{O}_{12}^{{\rm (S)}yzy}$,  $\mathcal{O}_{12}^{{\rm (S)}xxz}$,  $\mathcal{O}_{12}^{{\rm (S)}yyz}$, and  $\mathcal{O}_{12}^{{\rm (AS)}xyz}$ are induced, where the symmetry-allowed $\mathcal{O}_{12}^{{\rm (S)}zzz}$ is not induced in the present perturbation process.

Although the above results can be applied to spin systems with any spin length, it is noted in the case of a 
spin-half system; the electric quadrupole in Eq.~(\ref{eq:quadrupole}) is given by $Q^{\alpha\beta}_i=\delta_{\alpha\beta}/4$.
Then, the interaction in Eq.~(\ref{eq:HQ}) becomes
\begin{align} 
\mathcal{H}_\mathrm{Q}&= \frac{1}{4}
\sum_{\langle i,j\rangle}\sum_{\alpha,\gamma} \left\{ \mathcal{O}^{(\mathrm{S})\alpha\alpha\gamma}_{ij}(S^\gamma_j +S^\gamma_i) \right. \nonumber\\
 &\left.+\mathcal{O}^{(\mathrm{AS})\alpha\alpha\gamma}_{ij}(S^\gamma_j -S^\gamma_i)  \right\}, 
\end{align}
where the first (second) term corresponds to a uniform (staggered) magnetic field for two spins.

\begin{table*}
\caption{
\label{tab:class_OS}
Nonzero components in $\mathcal{O}^{(\mathrm{S})}_{12}$ under point group \textbf{G}.
Components listed here are induced by the light through $(Y^x_{12},Y^y_{12})$ shown in Table~\ref{tab:class_static}.
}
\begin{ruledtabular}
\begin{tabular}{cccc}
\textbf{G} & $\mathcal{O}^{(\mathrm{S})\alpha\beta x}_{12}$ & $\mathcal{O}^{(\mathrm{S})\alpha\beta y}_{12}$ & $\mathcal{O}^{(\mathrm{S})\alpha\beta z}_{12}$ \\ \hline
$mmm$ &
\\ 
$2mm$ & $\mathcal{O}^{(\mathrm{S})zx x}_{12}$
& $\mathcal{O}^{(\mathrm{S})yz y}_{12}$
& $\mathcal{O}^{(\mathrm{S})xx z}_{12}$, $\mathcal{O}^{(\mathrm{S})yy z}_{12}$
\\
$m2m$ & $\mathcal{O}^{(\mathrm{S})zx x}_{12}$
& $\mathcal{O}^{(\mathrm{S})yz y}_{12}$
& $\mathcal{O}^{(\mathrm{S})xx z}_{12}$, $\mathcal{O}^{(\mathrm{S})yy z}_{12}$ 
 \\
$mm2$ & $\mathcal{O}^{(\mathrm{S})zx x}_{12}$
& $\mathcal{O}^{(\mathrm{S})yz y}_{12}$
& $\mathcal{O}^{(\mathrm{S})zz z}_{12}$
\\
$222$ & $\mathcal{O}^{(\mathrm{S})zx x}_{12}$
& $\mathcal{O}^{(\mathrm{S})yz y}_{12}$
& $\mathcal{O}^{(\mathrm{S})zz z}_{12}$
\\
$2/m..$ & $\mathcal{O}^{(\mathrm{S})xy x}_{12}$,$\mathcal{O}^{(\mathrm{S})zx x}_{12}$
& $\mathcal{O}^{(\mathrm{S})yy y}_{12}$, $\mathcal{O}^{(\mathrm{S})yz y}_{12}$
& $\mathcal{O}^{(\mathrm{S})zz z}_{12}$, $\mathcal{O}^{(\mathrm{S})yz z}_{12}$
\\ 
$.2/m.$ & $\mathcal{O}^{(\mathrm{S})xx x}_{12}$, $\mathcal{O}^{(\mathrm{S})zx x}_{12}$
& $\mathcal{O}^{(\mathrm{S})yz y}_{12}$, $\mathcal{O}^{(\mathrm{S})xy y}_{12}$
& $\mathcal{O}^{(\mathrm{S})zz z}_{12}$, $\mathcal{O}^{(\mathrm{S})zx z}_{12}$
\\
$..2/m$ &
\\
$m..$ & $\mathcal{O}^{(\mathrm{S})xy x}_{12}$,$\mathcal{O}^{(\mathrm{S})zx x}_{12}$
& $\mathcal{O}^{(\mathrm{S})xx y}_{12}$, $\mathcal{O}^{(\mathrm{S})zz y}_{12}$$\mathcal{O}^{(\mathrm{S})yy y}_{12}$, $\mathcal{O}^{(\mathrm{S})yz y}_{12}$
& $\mathcal{O}^{(\mathrm{S})xx z}_{12}$, $\mathcal{O}^{(\mathrm{S})yy z}_{12}$, $\mathcal{O}^{(\mathrm{S})zz z}_{12}$, $\mathcal{O}^{(\mathrm{S})yz z}_{12}$ 
\\
$.m.$ & $\mathcal{O}^{(\mathrm{S})xx x}_{12}$, $\mathcal{O}^{(\mathrm{S})yy x}_{12}$, $\mathcal{O}^{(\mathrm{S})zz x}_{12}$, $\mathcal{O}^{(\mathrm{S})zx x}_{12}$
& $\mathcal{O}^{(\mathrm{S})yz y}_{12}$, $\mathcal{O}^{(\mathrm{S})xy y}_{12}$
& $\mathcal{O}^{(\mathrm{S})xx z}_{12}$, $\mathcal{O}^{(\mathrm{S})yy z}_{12}$, $\mathcal{O}^{(\mathrm{S})zz z}_{12}$, $\mathcal{O}^{(\mathrm{S})zx z}_{12}$
\\
$..m$ & $\mathcal{O}^{(\mathrm{S})yz x}_{12}$, $\mathcal{O}^{(\mathrm{S})zx x}_{12}$
& $\mathcal{O}^{(\mathrm{S})yz y}_{12}$, $\mathcal{O}^{(\mathrm{S})zx y}_{12}$
& $\mathcal{O}^{(\mathrm{S})xx z}_{12}$, $\mathcal{O}^{(\mathrm{S})yy z}_{12}$, $\mathcal{O}^{(\mathrm{S})xy z}_{12}$
\\
$2..$ & $\mathcal{O}^{(\mathrm{S})xy x}_{12}$,$\mathcal{O}^{(\mathrm{S})zx x}_{12}$
& $\mathcal{O}^{(\mathrm{S})xx y}_{12}$, $\mathcal{O}^{(\mathrm{S})zz y}_{12}$$\mathcal{O}^{(\mathrm{S})yy y}_{12}$, $\mathcal{O}^{(\mathrm{S})yz y}_{12}$
& $\mathcal{O}^{(\mathrm{S})xx z}_{12}$, $\mathcal{O}^{(\mathrm{S})yy z}_{12}$, $\mathcal{O}^{(\mathrm{S})zz z}_{12}$, $\mathcal{O}^{(\mathrm{S})yz z}_{12}$
\\
$.2.$ & $\mathcal{O}^{(\mathrm{S})xx x}_{12}$, $\mathcal{O}^{(\mathrm{S})yy x}_{12}$, $\mathcal{O}^{(\mathrm{S})zz x}_{12}$, $\mathcal{O}^{(\mathrm{S})zx x}_{12}$
& $\mathcal{O}^{(\mathrm{S})yz y}_{12}$, $\mathcal{O}^{(\mathrm{S})xy y}_{12}$
& $\mathcal{O}^{(\mathrm{S})xx z}_{12}$, $\mathcal{O}^{(\mathrm{S})yy z}_{12}$, $\mathcal{O}^{(\mathrm{S})zz z}_{12}$, $\mathcal{O}^{(\mathrm{S})zx z}_{12}$
\\
$..2$ & $\mathcal{O}^{(\mathrm{S})zx x}_{12}$
& $\mathcal{O}^{(\mathrm{S})yz y}_{12}$
& $\mathcal{O}^{(\mathrm{S})zz z}_{12}$
\\
$\bar{1}$ & $\mathcal{O}^{(\mathrm{S})xx x}_{12}$,$\mathcal{O}^{(\mathrm{S})xy x}_{12}$,$\mathcal{O}^{(\mathrm{S})zx x}_{12}$
& $\mathcal{O}^{(\mathrm{S})yy y}_{12}$, $\mathcal{O}^{(\mathrm{S})xy y}_{12}$,$\mathcal{O}^{(\mathrm{S})yz y}_{12}$
& $\mathcal{O}^{(\mathrm{S})zz z}_{12}$,$\mathcal{O}^{(\mathrm{S})yz z}_{12}$,$\mathcal{O}^{(\mathrm{S})zx z}_{12}$
\end{tabular}
\end{ruledtabular}
\end{table*}

\begin{table*}
\caption{
\label{tab:class_OAS}
Nonzero components in $\mathcal{O}^{(\mathrm{AS})}_{12}$ under point group \textbf{G}.
Components listed here are induced by the light through $(Y^x_{12},Y^y_{12})$ shown in Table~\ref{tab:class_static}.
}
\begin{ruledtabular}
\begin{tabular}{cccc}
\textbf{G} & $\mathcal{O}^{(\mathrm{AS})\alpha\beta x}_{12}$ & $\mathcal{O}^{(\mathrm{AS})\alpha\beta y}_{12}$ & $\mathcal{O}^{(\mathrm{AS})\alpha\beta z}_{12}$ \\ \hline
$mmm$ &
\\ 
$2mm$ & $\mathcal{O}^{(\mathrm{AS})zx x}_{12}$
& $\mathcal{O}^{(\mathrm{AS})yz y}_{12}$
& $\mathcal{O}^{(\mathrm{AS})xx z}_{12}$, $\mathcal{O}^{(\mathrm{AS})yy z}_{12}$
\\
$m2m$ &
&
& $\mathcal{O}^{(\mathrm{AS})xy z}_{12}$
 \\
$mm2$ & $\mathcal{O}^{(\mathrm{AS})xxx}_{12}$
& $\mathcal{O}^{(\mathrm{AS})xy y}_{12}$
& $\mathcal{O}^{(\mathrm{AS})zx z}_{12}$
\\
$222$ & $\mathcal{O}^{(\mathrm{AS})xy x}_{12}$
& $\mathcal{O}^{(\mathrm{AS})yy y}_{12}$
& $\mathcal{O}^{(\mathrm{AS})yz z}_{12}$
\\
$2/m..$ 
\\ 
$.2/m.$  
\\
$..2/m$ 
\\
$m..$ & $\mathcal{O}^{(\mathrm{AS})xx x}_{12}$, $\mathcal{O}^{(\mathrm{AS})yy x}_{12}$, $\mathcal{O}^{(\mathrm{AS})zz x}_{12}$, $\mathcal{O}^{(\mathrm{AS})yz x}_{12}$
& $\mathcal{O}^{(\mathrm{AS})xy y}_{12}$,  
& $\mathcal{O}^{(\mathrm{AS})xy z}_{12}$, $\mathcal{O}^{(\mathrm{AS})zx z}_{12}$
\\
$.m.$ & $\mathcal{O}^{(\mathrm{AS})xx x}_{12}$, $\mathcal{O}^{(\mathrm{AS})yy x}_{12}$, $\mathcal{O}^{(\mathrm{AS})zz x}_{12}$, $\mathcal{O}^{(\mathrm{AS})zx x}_{12}$ 
& $\mathcal{O}^{(\mathrm{AS})xy y}_{12}$, $\mathcal{O}^{(\mathrm{AS})yz y}_{12}$
& $\mathcal{O}^{(\mathrm{AS})xx z}_{12}$, $\mathcal{O}^{(\mathrm{AS})yy z}_{12}$, $\mathcal{O}^{(\mathrm{AS})zz z}_{12}$, $\mathcal{O}^{(\mathrm{AS})zx z}_{12}$
\\
$..m$ & $\mathcal{O}^{(\mathrm{AS})zx x}_{12}$
& $\mathcal{O}^{(\mathrm{AS})yz y}_{12}$
& $\mathcal{O}^{(\mathrm{AS})xx z}_{12}$, $\mathcal{O}^{(\mathrm{AS})yy z}_{12}$, $\mathcal{O}^{(\mathrm{AS})xy z}_{12}$
\\
$2..$ & $\mathcal{O}^{(\mathrm{AS})xy x}_{12}$, $\mathcal{O}^{(\mathrm{AS})zx x}_{12}$
& $\mathcal{O}^{(\mathrm{AS})xx y}_{12}$, $\mathcal{O}^{(\mathrm{AS})yy y}_{12}$, $\mathcal{O}^{(\mathrm{AS})zz y}_{12}$, $\mathcal{O}^{(\mathrm{AS})yz y}_{12}$
& $\mathcal{O}^{(\mathrm{AS})xx z}_{12}$, $\mathcal{O}^{(\mathrm{AS})yy z}_{12}$, $\mathcal{O}^{(\mathrm{AS})zz z}_{12}$, $\mathcal{O}^{(\mathrm{AS})yz z}_{12}$
\\
$.2.$ & $\mathcal{O}^{(\mathrm{AS})xy x}_{12}$,$\mathcal{O}^{(\mathrm{AS})yz x}_{12}$
& $\mathcal{O}^{(\mathrm{AS})xx y}_{12}$,  $\mathcal{O}^{(\mathrm{AS})yy y}_{12}$,$\mathcal{O}^{(\mathrm{AS})zz y}_{12}$,$\mathcal{O}^{(\mathrm{AS})zx y}_{12}$
&$\mathcal{O}^{(\mathrm{AS})xy z}_{12}$
, $\mathcal{O}^{(\mathrm{AS})yz z}_{12}$
\\
$..2$ & $\mathcal{O}^{(\mathrm{AS})xx x}_{12}$ $\mathcal{O}^{(\mathrm{AS})xy x}_{12}$
& $\mathcal{O}^{(\mathrm{AS})yy y}_{12}$, $\mathcal{O}^{(\mathrm{AS})xy y}_{12}$
& $\mathcal{O}^{(\mathrm{AS})yz z}_{12}$, $\mathcal{O}^{(\mathrm{AS})zx z}_{12}$
\\
$\bar{1}$ &
\end{tabular}
\end{ruledtabular}
\end{table*}

\begin{table*}
\caption{
\label{tab:class_M}
Symmetry rules of the two-site three-spin interactions: $\{\mathcal{O}^{(\mathrm{S/AS})xxx}_{12}\}=\{\mathcal{O}^{(\mathrm{S/AS})xxx}_{12},\mathcal{O}^{(\mathrm{S/AS})yyx}_{12},\mathcal{O}^{(\mathrm{S/AS})zzx}_{12},\mathcal{O}^{(\mathrm{S/AS})xyy}_{12},\mathcal{O}^{(\mathrm{S/AS})zxz}_{12}\}$,
 $\{\mathcal{O}^{(\mathrm{S/AS})yyy}_{12}\}=\{\mathcal{O}^{(\mathrm{S/AS})xyx}_{12},\mathcal{O}^{(\mathrm{S/AS})xxy}_{12},\mathcal{O}^{(\mathrm{S/AS})yyy}_{12},\mathcal{O}^{(\mathrm{S/AS})zzy}_{12},\mathcal{O}^{(\mathrm{S/AS})yzz}_{12}\}$,
 $\{\mathcal{O}^{(\mathrm{S/AS})zzz}_{12}\}=\{\mathcal{O}^{(\mathrm{S/AS})zxx}_{12},\mathcal{O}^{(\mathrm{S/AS})yzy}_{12},\mathcal{O}^{(\mathrm{S/AS})xxz}_{12},\mathcal{O}^{(\mathrm{S/AS})yyz}_{12},\mathcal{O}^{(\mathrm{S/AS})zzz}_{12}\}$,
and $\{\mathcal{O}^{(\mathrm{S/AS})xyz}_{12}\}=\{\mathcal{O}^{(\mathrm{S/AS})yzx}_{12},\mathcal{O}^{(\mathrm{S/AS})zxy}_{12},\mathcal{O}^{(\mathrm{S/AS})xyz}_{12}\}.$
All ($0$) means that all the components in $\{\cdots\}$ are symmetry-allowed (zero).
}
\begin{ruledtabular}
\begin{tabular}{ccccccccc}
\textbf{M} & $\{\mathcal{O}^{(\mathrm{S})xxx}_{12}\}$ & $\{\mathcal{O}^{(\mathrm{S})yyy}_{12}\}$ & $\{\mathcal{O}^{(\mathrm{S})zzz}_{12}\}$ & $\{\mathcal{O}^{(\mathrm{S})xyz}_{12}\}$ & $\{\mathcal{O}^{(\mathrm{AS})xxx}_{12}\}$ & $\{\mathcal{O}^{(\mathrm{AS})yyy}_{12}\}$ & $\{\mathcal{O}^{(\mathrm{AS})zzz}_{12}\}$ & $\{\mathcal{O}^{(\mathrm{AS})xyz}_{12}\}$ 
\\ \hline 
$m'm'm$ &$0$ &$0$ & All &$0$ 
&$0$ &$0$ &$0$ &$0$
\\ 
$2'm'm$ & $0$ & $0$ & All & $0$ 
& $0$ & $0$ & All & $0$
\\
$m'2'm$ & $0$ & $0$ & All &$0$ 
& $0$ & $0$ & $0$ & All
 \\
 $m'm'2$ &$0$ &$0$ & All & $0$ 
& All &$0$ &$0$ &$0$
\\
$2'2'2$ &$0$ &$0$ & All &$0$ 
&$0$ & All & $0$ &$0$
\\
$2'/m'..$ &$0$ & All & All & $0$
& $0$ & $0$ & $0$ & $0$
\\ 
$.2'/m'.$ & All & $0$ & All &$0$ 
& $0$ & $0$ & $0$ & $0$
\\
$..2/m$ & $0$ & $0$ & All & All 
& $0$ & $0$ & $0$ & $0$
\\
$m'..$ & $0$ & All & All & $0$
& All &$0$ & $0$ & All
\\
$.m'.$ & All & $0$ & All & $0$
& All & $0$ & All & $0$
 \\
$..m$ & $0$ & $0$ & All & All 
& $0$ & $0$ & All & All
\\
$2'..$ & $0$ & All & All & $0$ 
& $0$ & All & All & $0$
\\
$.2'.$ & All & $0$ & All & $0$ 
&$0$ & All & $0$ & All
\\
$..2$ & $0$ & $0$ & All & All 
& All & All & $0$ & $0$
\\ 
$\bar{1}$ & All & All & All & All
& $0$ & $0$ & $0$ & $0$
\end{tabular}
\end{ruledtabular}
\end{table*}

\subsection{Anisotropic three-site three-spin interaction}
\label{sec:3site}   

The three-site three-spin interaction is obtained from the commutation relation between the electric dipoles at different sites in $\mathcal{H}_\mathrm{3spin}$ as
\begin{align} 
 \mathcal{H}_\mathrm{3site}&= -\frac{i\delta E_0^2}{2\Omega}\sum_j\sum_{i\neq k} ([p^x_{ij},p^y_{jk}]+[p^x_{jk},p^y_{ij}])\nonumber\\
 &= \frac{\delta E_0^2}{2\Omega}\sum_j \sum_{i\neq k}\sum_{\alpha,\beta,\gamma,\zeta,\eta}\epsilon_{\beta\gamma\eta} \nonumber\\
 & \times (Y^{x;\alpha\beta}_{ij}Y^{y;\gamma\zeta}_{jk}-Y^{y;\alpha\beta}_{ij}Y^{x;\gamma\zeta}_{jk})S^\alpha_i S^\eta_j S^\zeta_k.
 \end{align}
Similarly to the anisotropic two-site two-spin and two-site three-spin interactions, nonzero components of the three-site three-spin interaction are determined by the crystal symmetry via the third-rank ME tensor.
Meanwhile, the symmetry conditions between them are different from each other: The anisotropic two-site two-spin and two-site three-spin interactions depend on the symmetry of the bond, while the three-site three-spin interactions depend on the symmetry of the plaquette consisting of the sites $i$, $j$, and $k$. 
Thus, the presence of the three-site three-spin interactions depend on the relative positions of the sites $i$, $j$, and $k$, which indicate that they are not simply classified by the point groups in Table.~\ref{tab:class_static}.
From the symmetry viewpoint, the emergence of the three-site three-spin interactions is explained by the change of the symmetry of the plaquette into the black and white magnetic point group, as shown in Sec.~\ref{sec:H3site_tri}.

\section{Application to a triangular unit }
\label{sec:application}

\begin{figure}[t!]
\begin{center}
\includegraphics[width=1.0\hsize]{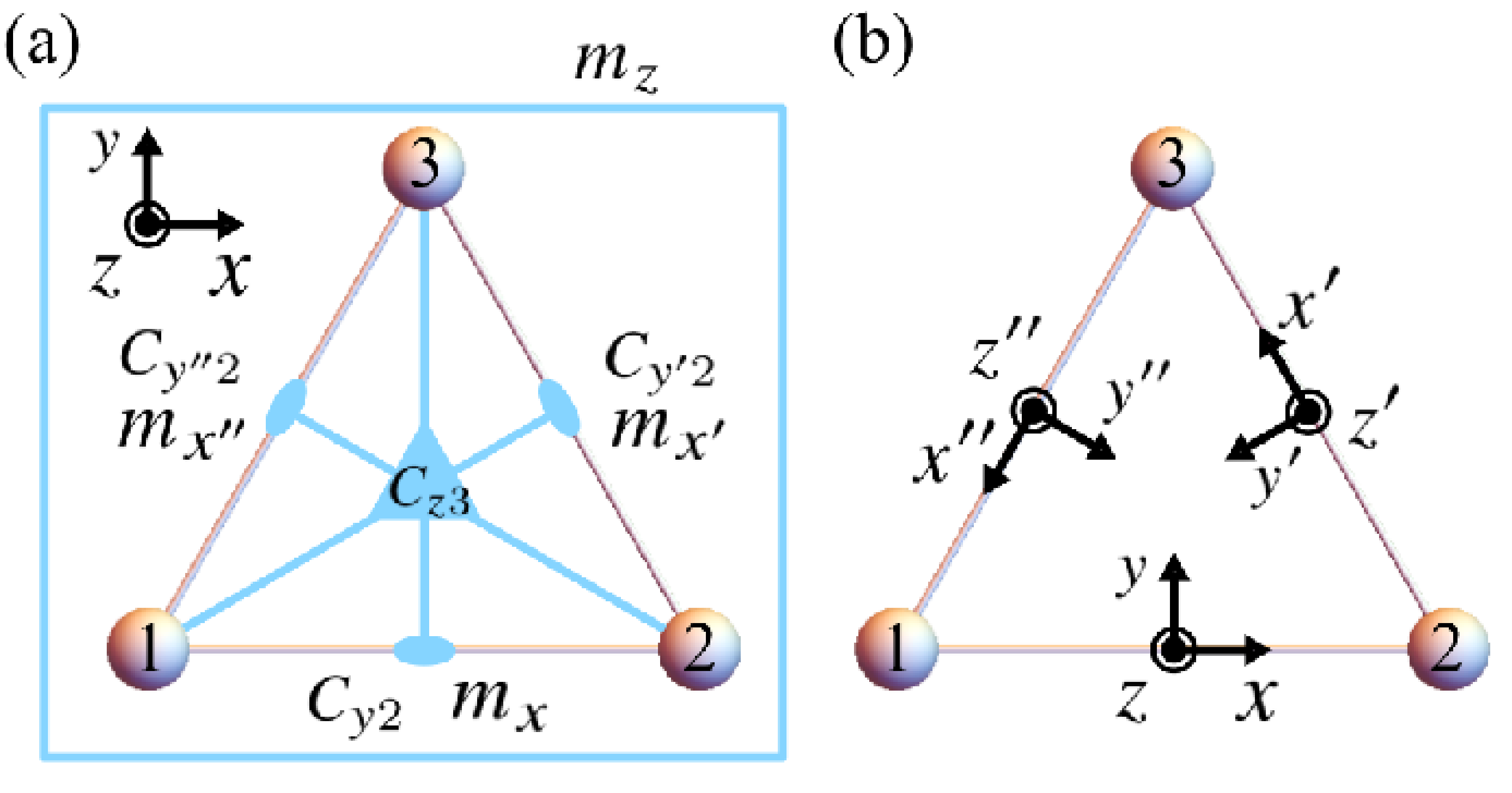} 
\caption{\label{fig:triangle}
(a) Triangular unit under the point group $m2\bar{6}$.
(b) Local Cartesian spin coordinates $(x,y,z)$ for the $\langle 1,2 \rangle$ bond, $(x',y',z')$ for the $\langle 2,3 \rangle$ bond, and $(x'',y'',z'')$ for the $\langle 3,1 \rangle$ bond.
}
\end{center}
\end{figure}
\begin{figure}[t!]
\begin{center}
\includegraphics[width=1.0\hsize]{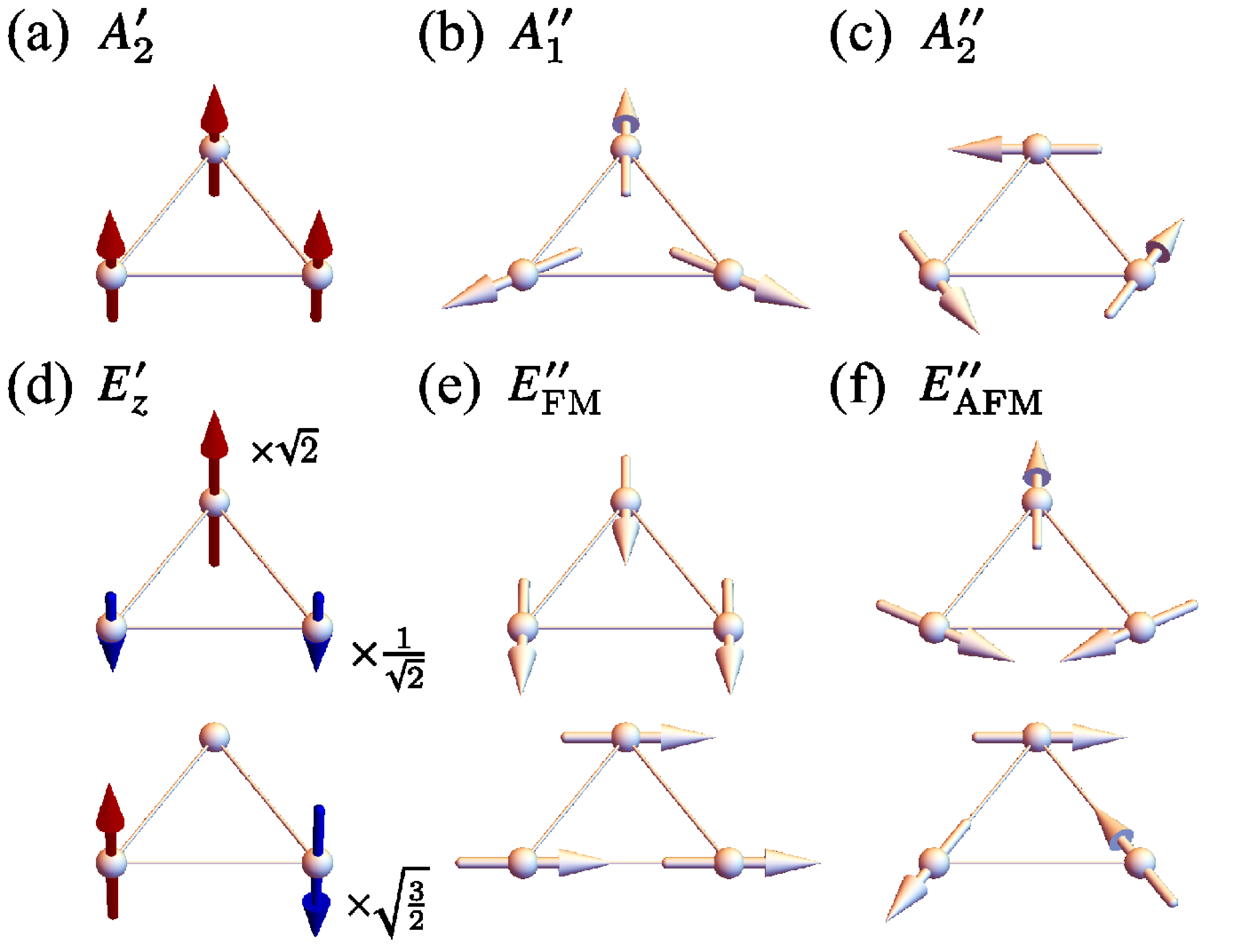} 
\caption{
\label{fig:irrep}
Bases for the irreducible representations (a) $A'_2$, (b) $A''_1$, (c) $A''_2$, (d) $E'_z$, (e) $E''_{\rm FM}$, and (f) $E''_{\rm AFM}$.
Upper and lower panels in two-dimensional representations (d)-(f) correspond 
to the bases with the same representation matrix as that for $x$ and $y$, respectively.
The color shows the $z$ spin components, where red, white, and blue correspond to the positive, zero, and negative ones, respectively.
The spin length at each site is fixed as $|\bm{S}_i|=1$ except for (d): $(|\bm{S}_1|,|\bm{S}_2|,|\bm{S}_3|)=(1/\sqrt{2},1/\sqrt{2},\sqrt{2})$ for the upper panel in (d) and $(|\bm{S}_1|,|\bm{S}_2|,|\bm{S}_3|)=(\sqrt{3/2},\sqrt{3/2},0)$ for the lower panel in (d).
The details of the spin configurations are shown in Appendix~\ref{app:bases}. 
}
\end{center}
\end{figure}

We apply the above results to a triangular unit with the point group $m2\bar{6}$ shown in Fig.~\ref{fig:triangle}.
By starting from a static Hamiltonian in Sec.~\ref{sec:H0_tri}, we show the light-induced one spin, anisotropic two-site two-spin, anisotropic two-site three-spin, anisotropic three-site three-spin interactions in Secs.~\ref{sec:H1spin_tri}-\ref{sec:H3site_tri}, respectively. 
In each case, we discuss the modulation of a spin structure under the light-induced anisotropic interactions by taking the classical spin limit for simplicity.

\subsection{Static Hamiltonian}
\label{sec:H0_tri}

We start from the static Hamiltonian in Eq.~(\ref{eq:H0}) in the $m2\bar{6}$ triangular unit, which is given by
\begin{align}
\mathcal{H}_0&=\sum_{\alpha,\beta}\left( 
J^{\alpha\beta}_{12}S^\alpha_1S^\beta_2
+J^{\alpha\beta}_{23}S^\alpha_2S^\beta_3
+J^{\alpha\beta}_{31}S^\alpha_3S^\beta_1
\right) \nonumber\\
\label{eq:H0_tri}
&=\sum_{\alpha,\beta}J^{\alpha\beta}_{12}\left( 
S^\alpha_1S^\beta_2
+S^{\alpha'}_2S^{\beta'}_3
+S^{\alpha''}_3S^{\beta''}_1
\right).
\end{align}
The Hamiltonian in the first line is written in the global Cartesian spin coordinate $\bm{S}=(S^{x},S^{y},S^{z})$.
We fix the spin length $|\bm{S}_i|=1$ in each site for simplicity.
The coupling matrices $J^{\alpha\beta}_{23}$, and $J^{\alpha\beta}_{31}$ are related to $J^{\alpha\beta}_{12}$ due to the threefold rotation around the $z$ axis. 
By using the local Cartesian spin coordinates $\bm{S}$ for the $\langle 1,2 \rangle$ bond, $\bm{S}'=(S^{x'},S^{y'},S^{z'})$ for the $\langle 2,3 \rangle$ bond, and $\bm{S}''=(S^{x''},S^{y''},S^{z''})$ for the $\langle 3,1 \rangle$ bond shown in Fig.~\ref{fig:triangle}(b), the Hamiltonian is represented by the second line. 
Thus, the Hamiltonian is characterized by $J_{12}$, which is given by
\begin{align}
\label{eq:J}
J_{12} &= \begin{pmatrix}
F^x & D^z & 0 \\
-D^z & F^y & 0 \\
0 & 0 & F^z 
\end{pmatrix}.
\end{align}
Nonzero components in $J_{12}$ are determined by the point group $m2m$ of the $\langle 1,2 \rangle$ bond (see also Table~\ref{tab:class_static}). 
$J^{\alpha\beta}_{23}$ and $J^{\alpha\beta}_{31}$ written in the global Cartesian spin coordinate are shown in Appendix~\ref{app:matrix}.

\begin{table}
\caption{
\label{tab:character}
Irreducible representations and character table for the point group $m2\bar{6}$.
}
\begin{ruledtabular}
\begin{tabular}{ccccccc}
 &$E$ & $m_z$ & $2C_{z3}$ & $2S_6$ & $3C_{y2}$ & $3m_x$ \\ \hline
$A'_1$ & 1 & 1 & 1 & 1 & 1 & 1 \\
$A'_2$ & 1 & 1 & 1 & 1 & -1 & -1 \\
$A''_1$ & 1 & -1 & 1 & -1 & 1 & -1 \\
$A''_2$ & 1 & -1 & 1 & -1 & -1 & 1 \\
$E'$ & 2 & 2 & -1 & -1 & 0 & 0 \\
$E''$ & 2 & -2 & -1 & 1 & 0 & 0 
\end{tabular}
\end{ruledtabular}
\end{table}

We discuss the ground-state spin configuration in $\mathcal{H}_{0}$ by using the irreducible representation $\Gamma$ under the point group $m2\bar{6}$: $\Gamma=A'_1$, $A'_2$, $A''_1$, $A''_2$, $E'$, and $E''$ shown in Table~\ref{tab:character}. 
The arbitrary spin configuration $S^\triangle=(\bm{S}_1,\bm{S}_2,\bm{S}_3)$ is expressed by bases $S(\Gamma)$ for the irreducible representation $\Gamma$ as  
\begin{align}
\label{eq: Sirrep}
\tilde{S}&=m_{A'_2}S(A'_2)+m_{A''_1}S(A''_1)+m_{A''_2}S(A''_2)  \nonumber\\
&+\bm{m}_{E'_{z}}\cdot\bm{S}(E'_{z})+\bm{m}_{E''_{\mathrm{FM}}}\cdot\bm{S}(E''_{\mathrm{FM}}) \nonumber\\
&+\bm{m}_{E''_{\mathrm{AFM}}}\cdot\bm{S}(E''_{\mathrm{AFM}}),
\end{align}
where $S(\Gamma)$ is the nine-dimensional vector, whose component is given by the spin configuration shown in Fig.~\ref{fig:irrep}, and $m_{\Gamma}$ is the order parameter of $S(\Gamma)$.
We use $\bm{S}(\Gamma)=(S^x_{\Gamma},S^y_{\Gamma})$ and $\bm{m}_{\Gamma}=(m^x_{\Gamma},m^y_{\Gamma})$ for the two-dimensional representations.
By using Eq.~(\ref{eq: Sirrep}), the Hamiltonian in Eq.~(\ref{eq:H0_tri}) is rewritten as
\begin{align}
\mathcal{H}_0&=\frac{3}{2}(\lambda_{A'_2}m_{A'_2}^2+\lambda_{A''_1}m_{A''_1}^2+\lambda_{A''_2}m_{A''_2}^2  \nonumber\\
&+\lambda_{E'_z}\bm{m}_{E'_{z}}^2+\lambda_{E''_{\mathrm{FM}}}\bm{m}_{E''_{\mathrm{FM}}}^2+\lambda_{E''_{\mathrm{AFM}}}\bm{m}_{E''_{\mathrm{AFM}}}^2 \nonumber \\
&+\lambda_{E''} \bm{m}_{E''_{\mathrm{FM}}}\cdot\bm{m}_{E''_{\mathrm{AFM}}}),
\end{align}
with
\begin{align}
\lambda_{A'_2}&=2F^z,  \\
\lambda_{A''_1}&=\frac{1}{2}(-3F^x+F^y+2\sqrt{3}D^z), \\
\lambda_{A''_2}&=\frac{1}{2}(F^x-3F^y+2\sqrt{3}D^z), \\
\lambda_{E'_z}&=-F^z, \\
\lambda_{E''_{\mathrm{FM}}}&=F^x+F^y,\\
\lambda_{E''_{\mathrm{AFM}}}&=\frac{1}{2}(-F^x-F^y-2\sqrt{3}D^z), \\
\lambda_{E''}&=-F^x+F^y.
\end{align}
The ground-state spin configuration in the classical spin limit is obtained as follows.
Due to constraint on the spin length at each site ($|\bm{S}_i|=1$), the order parameters satisfy  $|m^\alpha_\Gamma|\le 1$ and  $\sum_{\alpha,\Gamma}(m^\alpha_\Gamma)^2=1$.  
Then, the ground-state spin configuration is given by $\bm{S}^\triangle$ with $|m_{\Gamma_{\rm min}}|=1$ and $m_{\Gamma\neq\Gamma_{\rm min}}=0$, where the state with $\Gamma_{\rm min}$ gives the smallest eigenvalue $\lambda_{\Gamma_{\rm min}}$.
Various spin configurations become the ground state depending on the model parameters in $J_{12}$~\cite{PhysRevB.96.205126}, while all of them are collinear, $S(A'_2)$ and $\bm{S}(E'_z)$, or coplanar, $S(A''_1)$, $S(A''_2)$, $\bm{S}(E''_\mathrm{FM})$, and $\bm{S}(E''_\mathrm{AFM})$.
This is because the coupling matrix $J_{12}$ has no components mixing the $z$ and $xy$ spins, such as  $J^{yz}_{12}$ and $J^{zx}_{12}$, due to the horizontal mirror symmetry $m_z$.    

\subsection{Light-induced magnetic interactions}

We show the light-induced magnetic interactions in the triangular unit and discuss a modulation of the spin configuration in the classical spin limit.  
The third-rank ME tensor on the $\langle 1,2 \rangle$ bond is given by
\begin{align}
Y^x_{12}&=
\begin{pmatrix}
0 & B_{12}^{x;z} & 0 \\
B_{12}^{x;z} & 0 & 0\\
0 & 0 & 0
\end{pmatrix},\\
Y^y_{12}&=
\begin{pmatrix}
C_{12}^{y;x} & A_{12}^{y;z} & 0 \\
-A_{12}^{y;z} & C_{12}^{y;y} & 0\\
0 & 0 & C_{12}^{y;z}
\end{pmatrix}.
\end{align}
Here, nonzero components are determined by the point group symmetry $m2m$ of the $\langle 1,2 \rangle$ bond, as shown in Table.~\ref{tab:class_static}.
The third-rank ME tensors on the other bonds are given by components in $(Y^x_{12},Y^y_{12})$ via the threefold rotation, as shown in Appendix~\ref{app:matrix}. 

\subsubsection{$\mathcal{H}_{\rm 1spin}$}
\label{sec:H1spin_tri}

The light-induced one-spin term is given by
\begin{align}
\mathcal{H}_{\rm 1spin}&=\frac{\delta B_0^2}{2\Omega}(S^z_1+S^z_2+S^z_3) \nonumber\\
&=\frac{3\delta B_0^2}{2\Omega}m_{A'_2}.
\end{align}
Thus, $\mathcal{H}_{\rm 1spin}$ favors $S^\triangle=m_{A'_2}S(A'_2)$ with $m_{A'_2}=-1$ ($m_{A'_2}=+1$) for the right-circularly (left-circularly) polarized light with $\delta=+1$ ($\delta=-1$).

\subsubsection{$\mathcal{H}_{\rm 2spin}$}
\label{sec:H2spin_tri}

\begin{figure}[t!]
\begin{center}
\includegraphics[width=1.0\hsize]{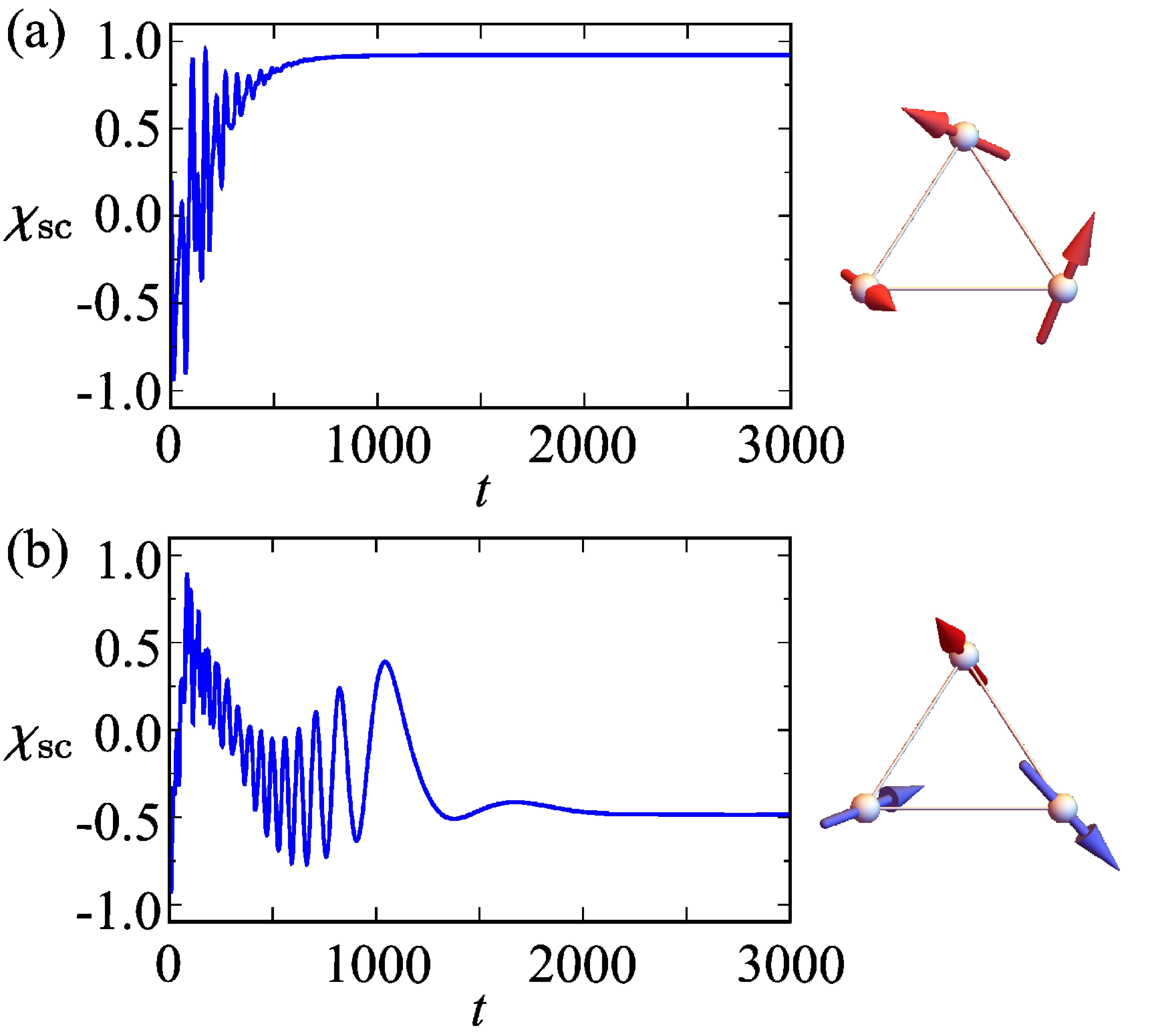} 
\caption{\label{fig:2spin}
Time evolutions of the spin scalar chirality $\chi_{\rm sc}$ by (a) $\mathcal{D}^x_{12}=1$ and (b) $\mathcal{E}^y_{12}=1$.
Right panels show the stable spin configurations in each case.
}
\end{center}
\end{figure}

In $\mathcal{H}_{\rm 2spin}$, we obtain the light-induced DM interaction $\mathcal{H}^{(1)}_{\rm 2spin}$ with $\mathcal{D}^x_{12}$ and the light-induced symmetric off-diagonal interaction $\mathcal{H}^{(2)}_{\rm 2spin}$ with $\mathcal{E}^y_{12}$.
Here, $\mathcal{D}^x_{12}$ and $\mathcal{E}^y_{12}$ are allowed by the chiral point group $\textbf{G}^{\rm(C)}=.2.$ of the $\langle 1,2\rangle$ bond. 
Reflecting the threefold rotation, the symmetry-equivalent interactions are induced by the light on the $\langle 2,3\rangle$ and $\langle 3,1\rangle$ bonds.

The light-induced DM interaction is given by
\begin{align}
\mathcal{H}^{(1)}_{\rm 2spin}&=\mathcal{D}^x_{12}(\bm{S}_1\times\bm{S}_2+\bm{S}'_2\times\bm{S}'_3+\bm{S}''_3\times\bm{S}''_1)^x \nonumber\\
&=-3\sqrt{3}\mathcal{D}^x_{12}\left[ m_{A'_2}m_{A''_2}+\frac{1}{\sqrt{2}}(\bm{m}_{E'_z}\times\bm{m}_{E''_\mathrm{FM}})^z \right],
\end{align}
where $\mathcal{D}^x_{12}=-\delta E_0B_0A^{y;z}_{12}/2\Omega$.
From the second line, we find that the DM interaction is represented by a linear combination of the terms belonging to $A''_1$ owing to the breaking of the mirror symmetries $\{m_x,m_y,m_z\}$ by the light.
In other words, the light-induced DM interaction belongs to the totally symmetric irreducible representation of the chiral point group $\textbf{G}^{\rm(C)}=.23$ of the triangle unit. 
In the classical spin limit, the DM interaction favors 
\begin{align}
S^\triangle&=m_{A'_2}S(A'_2)+m_{A''_2}S(A''_2),
\end{align}
with $m_{A'_2}=m_{A''_2}=\pm\frac{1}{\sqrt{2}}$ for $\mathcal{D}^x_{12}>0$ and $m_{A'_2}=-m_{A''_2}=\pm\frac{1}{\sqrt{2}}$ for $\mathcal{D}^x_{12}<0$, where the sign of $\mathcal{D}^x_{12}$ can be changed by the polarization $\delta$.
Thus, the light-induced interaction favors the spin configuration consisting of the superposition of the collinear configuration along the $z$ axis and the coplanar configuration on the $xy$ plane, which results in the noncoplanar spin configuration with the spin scalar chirality $\chi_{\rm sc}=\bm{S}_1\cdot\bm{S}_2\times\bm{S}_3$.
Meanwhile, the sign of $\chi_{\rm sc}$ is not determined by $\mathcal{D}^x_{12}$. 
It is noted that the same light-induced magnetic interactions are also obtained when directly starting from the classical spin
model instead of the quantum spin model in Eq.~(\ref{eq:Ht}) once the Gilbert damping is negligibly small~\cite{PhysRevLett.128.037201,higashikawa2018floquet}.

The light-induced symmetric off-diagonal interaction is given by
\begin{align}
\mathcal{H}^{(2)}_{\rm 2spin}&=\mathcal{E}^y_{12} \left[ (S^z_1S^x_2+S^x_1S^z_2)+(S^{z'}_2S^{x'}_3+S^{x'}_2S^{z'}_3) \right. \nonumber\\
&\left.+(S^{z''}_3S^{x''}_1+S^{x''}_3S^{z''}_1) \right] \\
&=3\mathcal{E}^y_{12}\left[ m_{A'_2}m_{A''_2} \right. \nonumber\\
&\left.+\frac{1}{\sqrt{2}}\{\bm{m}_{E'_z}\times (2\bm{m}_{E''_\mathrm{AFM}}-\bm{m}_{E''_\mathrm{FM}})\}^z\right],
\end{align}
where  $\mathcal{E}^{y}_{12}=\delta E_0B_0(-C^{y;z}_{12}+C^{y;x}_{12}-B^{x;z}_{12})/2\Omega$.
This interaction belongs to the irreducible representation $A''_1$ in the point group $m2\bar{6}$ or the totally symmetric irreducible representation in the chiral point group $.23$ as well as the light-induced DM interaction.
In the classical spin limit, the symmetric off-diagonal interaction mixes the colinear and coplanar configurations as
\begin{align}
S^\triangle &= \frac{\pm1}{\sqrt{10}}\left[ m^x_{E'_{\rm z}}S^x(E'_z)+ m^y_{E''_{\rm AFM}}S^y(E''_{\rm AFM})\right. \nonumber\\
&\left. +m^y_{E''_{\rm FM}}S^y(E''_{\rm FM}) \right]
\end{align}
where the positive and negative $\mathcal{E}^{y}_{12}$ favors $S^\triangle$ with  $(m^x_{E'_{\rm z}},m^y_{E''_{\rm AFM}},m^y_{E''_{\rm FM}})=(\sqrt{5},-2,1)/\sqrt{10}$ and $(m^x_{E'_{\rm z}},m^y_{E''_{\rm AFM}},m^y_{E''_{\rm FM}})=(\sqrt{5},2,-1)/\sqrt{10}$, respectively. 
This spin configuration accompanies nonzero spin scalar chirality, while the sign of $\chi_{\rm sc}$ is not determined.

We confirm the above analytical results by using numerical simulation.
By starting a random spin configuration, we calculate a time evolution of the spin scalar chirality $\chi_{\rm sc}$ by numerically solving the LLG equation in the classical spin limit ($|\bm{S}_i|=1$).
We analyze the Hamiltonian in the LLG equation including only the light-induced DM interaction $\mathcal{H}^{(1)}_{\rm 2spin}$ or symmetric off-diagonal interaction $\mathcal{H}^{(2)}_{\rm 2spin}$ to focus on their effect.
The time evolution of $\chi_{\rm sc}$ and stable spin configurations  by $\mathcal{D}^x_{12}=1$ and 
$\mathcal{E}^{y}_{12}=1$ are shown in Figs.~\ref{fig:2spin}(a) and \ref{fig:2spin}(b), respectively.  
These numerical results are consistent with the analytical results based on the irreducible representation. 
Although we obtain the spin configurations with $\chi_\mathrm{sc}>0$ ($\chi_\mathrm{sc}<0$) by $\mathcal{D}^x_{12}$ ($\mathcal{E}^y_{12}$), we also obtain the spin configuration with the opposite $\chi_\mathrm{sc}$ by starting from different random spin configurations.
It is noted that the stable spin configuration in Fig.~\ref{fig:2spin}(b) has small contributions from bases other than $S^x(E'_z)$, $S^y(E''_{\rm AFM})$, and $S^y(E''_{\rm FM})$ to satisfy the fixed-spin-length condition at each site.

\subsubsection{$\mathcal{H}_{\rm Q}$}
\label{sec:HQ_tri}

\begin{figure}[t!]
\begin{center}
\includegraphics[width=1.0\hsize]{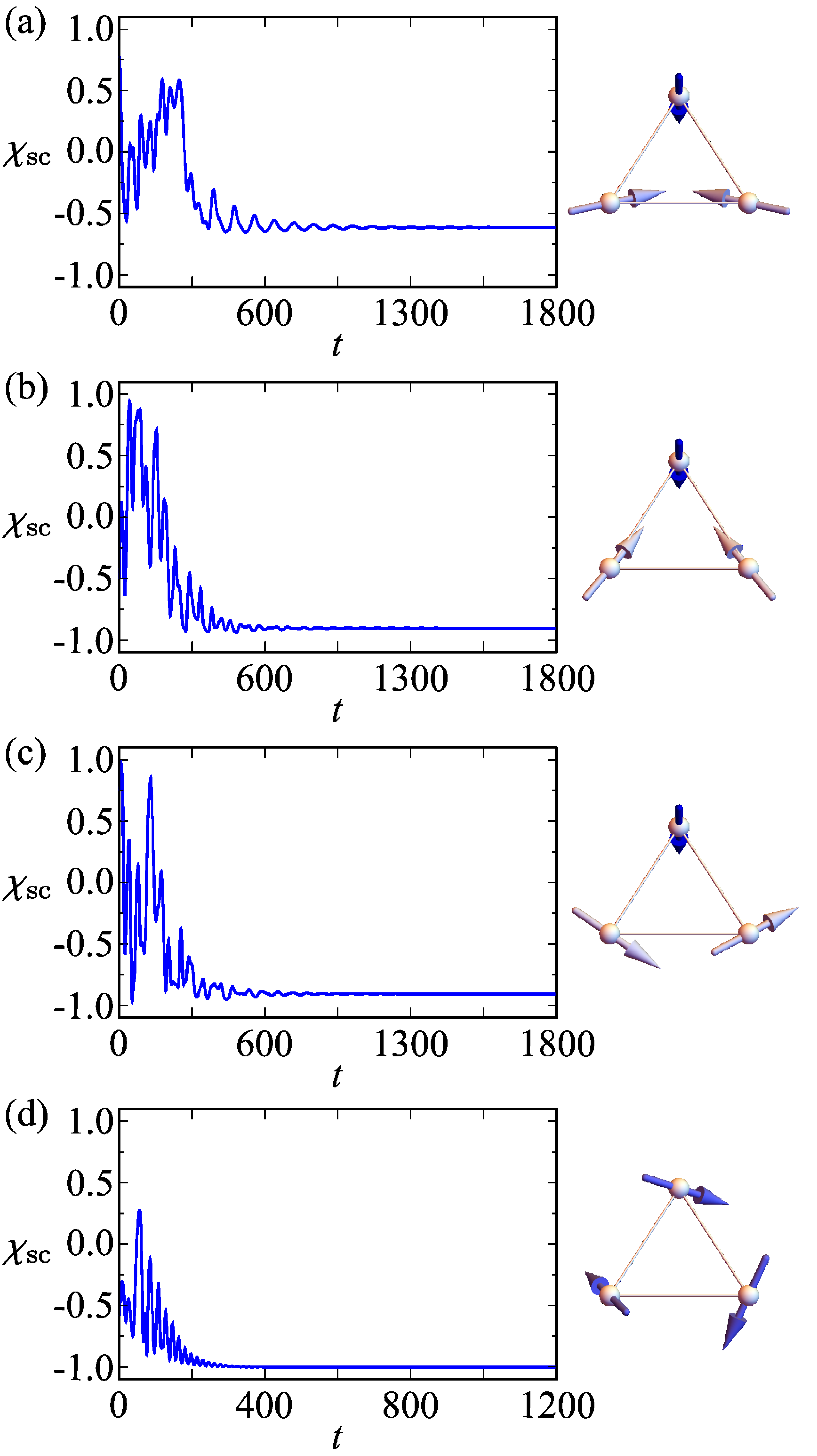} 
\caption{\label{fig:Q}
Time evolutions of the spin scalar chirality $\chi_{\rm sc}$ by (a) $\mathcal{O}^{\mathrm{(AS)}xyz}_{12}=1$, (b) $\mathcal{O}^{\mathrm{(S)}xxz}_{12}=1$, (c) $\mathcal{O}^{\mathrm{(S)}yyz}_{12}=1$, and (d) $\mathcal{O}^{\mathrm{(S)}wz}_{12}=1$.
Right panels show the stable spin configurations in each case.
}
\end{center}
\end{figure}

We obtain four types of the light-induced two-site three-spin interaction from Tables~\ref{tab:class_OS} and \ref{tab:class_OAS} as
\begin{align}
\mathcal{H}^{(1)}_{\rm Q}&=\mathcal{O}^{\mathrm{(AS)}xyz}_{12}\left[(Q_1^{xy}S_2^{z}-S_1^{z}Q_2^{xy})+(Q_2^{x'y'}S_3^{z'}-S_2^{z'}Q_3^{x'y'})\right. \nonumber \\
&\left.+(Q_3^{x''y''}S_1^{z''}-S_3^{z''}Q_1^{x''y''})\right],\\
\mathcal{H}^{(2)}_{\rm Q}&=\mathcal{O}^{\mathrm{(S)}xxz}_{12}\left[(Q_1^{xx}S_2^{z}+S_1^{z}Q_2^{xx})+(Q_2^{x'x'}S_3^{z'}+S_2^{z'}Q_3^{x'x'})\right. \nonumber \\
&\left.+(Q_3^{x''x''}S_1^{z''}+S_3^{z''}Q_1^{x''x''})\right],\\
\mathcal{H}^{(3)}_{\rm Q}&=\mathcal{O}^{\mathrm{(S)}yyz}_{12}\left[(Q_1^{yy}S_2^{z}+S_1^{z}Q_2^{yy})+(Q_2^{y'y'}S_3^{z'}+S_2^{z'}Q_3^{y'y'})\right. \nonumber \\
&\left.+(Q_3^{y''y''}S_1^{z''}+S_3^{z''}Q_1^{y''y''})\right],\\
\mathcal{H}^{(4)}_{\rm Q}&=\mathcal{O}^{\mathrm{(S)}wz}_{12}\left[(Q_1^{zx}S_2^{x}+S_1^{x}Q_2^{zx})+(Q_2^{z'x'}S_3^{x'}+S_2^{x'}Q_3^{z'x'})\right. \nonumber \\
&\left.+(Q_3^{z''x''}S_1^{x''}+S_3^{x''}Q_1^{z''x''})\right] \nonumber \\
&-\mathcal{O}^{\mathrm{(S)}wz}_{12}\left[(Q_1^{yz}S_2^{y}+S_1^{y}Q_2^{zy})+(Q_2^{y'z'}S_3^{y'}+S_2^{y'}Q_3^{z'y'})\right. \nonumber \\
&\left.+(Q_3^{y''z''}S_1^{y''}+S_3^{y''}Q_1^{z''y''})\right].
\end{align}
where $\mathcal{O}^{\mathrm{(AS)}xyz}_{12}=\delta E_0^2 B^{x;z}_{12}A^{y;z}_{12}/\Omega$, $\mathcal{O}^{\mathrm{(S)}xxz}_{12}=-\delta E_0^2 B^{x;z}_{12}C^{y;x}_{12}/2 \Omega$, $\mathcal{O}^{\mathrm{(S)}yyz}_{12}=\delta E_0^2 B^{x;z}_{12}C^{y;y}_{12}/2 \Omega$, and $\mathcal{O}^{\mathrm{(S)}wz}_{12}=\mathcal{O}^{\mathrm{(S)}zxx}_{12}=-\mathcal{O}^{\mathrm{(S)}yzy}_{12}=\delta E_0^2 B^{x;z}_{12}C^{y;z}_{12}/2\Omega$.
As shown in Sec.~\ref{sec:Q}, the light-induced two-site three-spin interaction on the $\langle 1,2 \rangle$ bond results from the breakings of the twofold rotation $C_{y2}$ and $m_x$ by light.
Meanwhile, the triangle system is invariant under the threefold rotation around the $z$ axis.
Accordingly, the obtained two-site three-spin interactions belong to the irreducible representation $A'_2$ with the odd parity for $\{C_{y2},m_x\}$ and the even parity for $C_{z3}$, as shown in Table~\ref{tab:character}.
It is noted that these interactions on the $\langle 1,2 \rangle$ bond (the triangle) belongs to the totally symmetric irreducible representation of the black and white magnetic point group $\textbf{M}=m'2'm$ ($\textbf{M}=m'2'\bar{6}$). 
These interactions can be expressed in the ternary of the order parameters, while it is cumbersome to analytically obtain the stable spin configuration.

We directly investigate stable spin configurations by numerically solving the LLG equation in the same manner in Sec.~\ref{sec:H2spin_tri}.
The time evolutions of $\chi_{\rm sc}$ and stable spin configurations by setting $\mathcal{O}^{\mathrm{(AS)}xyz}_{12}=1$, $\mathcal{O}^{\mathrm{(S)}xxz}_{12}=1$, $\mathcal{O}^{\mathrm{(S)}yyz}_{12}=1$, or $\mathcal{O}^{\mathrm{(S)}wz}_{12}=1$ are shown in Figs.~\ref{fig:Q}(a)-\ref{fig:Q}(d), respectively.  
The results show that all the two-site three-spin interactions favor noncoplanar spin configurations with nonzero spin scalar chirality. 
This is because the spin scalar chirality also belongs to $A'_2$ as well as the two-site three-spin interactions.
The mechanism of nonzero $\chi_\mathrm{sc}$ is different from the light-induced two-site two-spin interaction; the two-site two-spin interaction mixes the $z$ spin and the $xy$ spin as a result of the breaking of the horizontal mirror symmetry, while the two-site three-spin interaction is directly coupled to the spin scalar chirality belonging to the same representation.   
It is noted that the sign of $\chi_\mathrm{sc}$ is not determined by these interactions.

\subsubsection{$\mathcal{H}_{\rm 3site}$}
\label{sec:H3site_tri}

\begin{figure}[t!]
\begin{center}
\includegraphics[width=1.0\hsize]{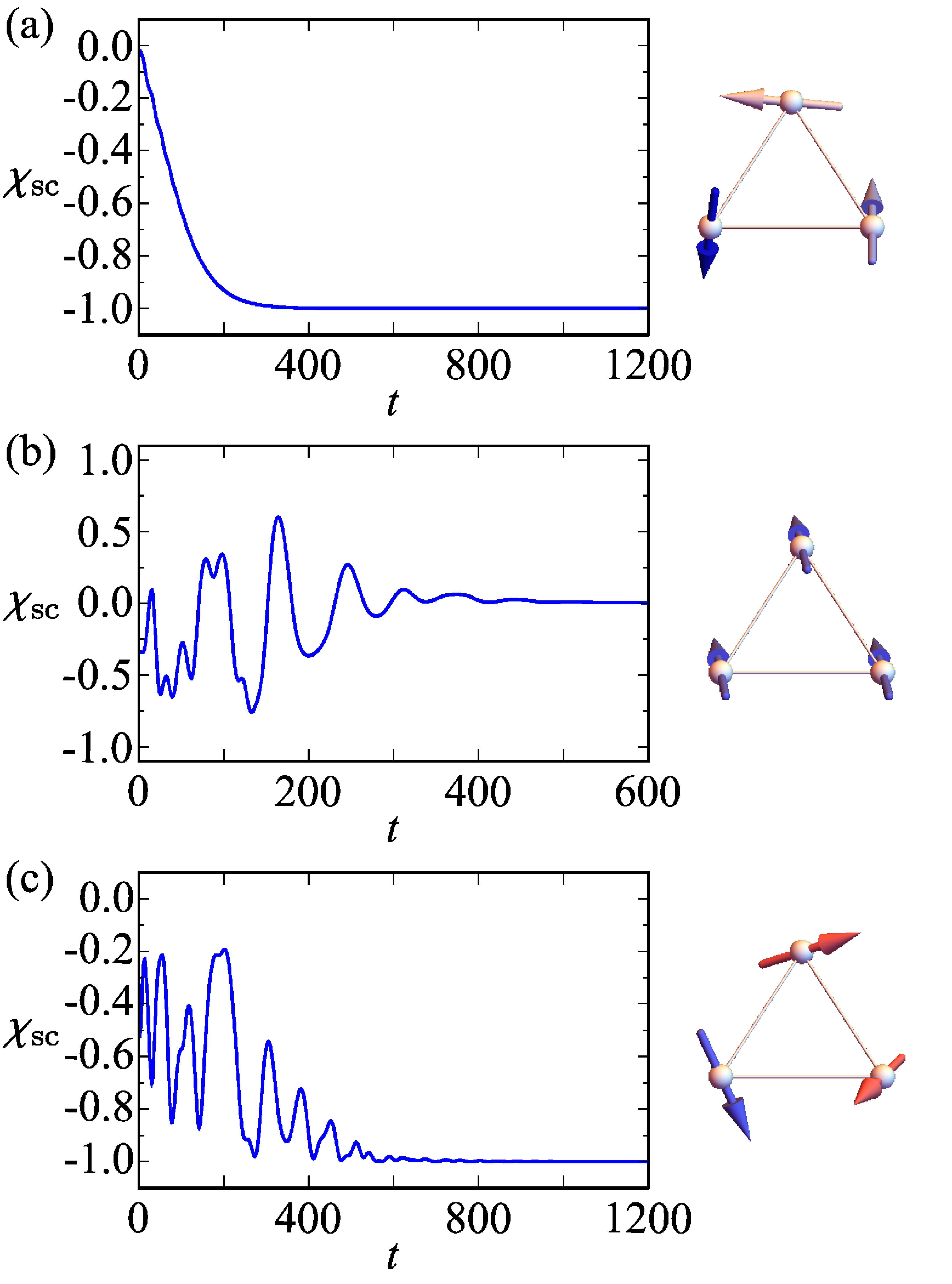} 
\caption{\label{fig:3site}
Time evolutions of the spin scalar chirality $\chi_{\rm sc}$ by (a) $\mathcal{T}^{\mathrm{(1)}}=1$, (b) $\mathcal{T}^{\mathrm{(2)}}=1$, and (c) $\mathcal{T}^{\mathrm{(3)}}=1$.
Right panels show the stable spin configurations in each case.
}
\end{center}
\end{figure}

Next, let us discuss the three-site three-spin interaction under the point group $\textbf{G}=m2\bar{6}$, where we obtain three types of the interactions as a consequence of the symmetry reduction from the point group $\textbf{G}=m2\bar{6}$ to the black and white magnetic point group $\textbf{M}=m'2'\bar{6}$. 
The obtained three-site spin interactions are given by
\begin{align}
\label{eq:H3site_1}
\mathcal{H}^{(1)}_{\rm 3site}&=\mathcal{T}^{(1)}\bm{S}_1\cdot\bm{S}_2\times\bm{S}_3,\\
\label{eq:H3site_2}
\mathcal{H}^{(2)}_{\rm 3site}&=\mathcal{T}^{(2)}\left[
S^z_1(S^x_2S^x_3+S^y_2S^y_3)
+S^z_2(S^x_3S^x_1+S^y_3S^y_1)\right. \nonumber\\
&\left.+S^z_3(S^x_1S^x_2+S^y_1S^y_2)
\right],\\
\label{eq:H3site_3}
\mathcal{H}^{(3)}_{\rm 3site}&=\mathcal{T}^{(3)}\left[
S^{z'}_1(S^{x'}_2S^{x'}_3-S^{y'}_2S^{y'}_3)\right. \nonumber\\
&\left.+S^{z'}_2(S^{x''}_3S^{x''}_1-S^{y''}_3S^{y''}_1)
+S^{z}_3(S^{x}_1S^{x}_2-S^{y}_2S^{y}_2)
\right],
\end{align}
with
\begin{align}
\mathcal{T}^{(1)}&=\frac{\sqrt{3}}{2}(A_{12}^{y;z})^2-\frac{\sqrt{3}}{4}(B_{12}^{x;z})^2-\frac{\sqrt{3}}{2}(C_{12}^{y;u})^2 \nonumber\\
&-\sqrt{3}C_{12}^{y;u}C_{12}^{y;z}+\frac{\sqrt{3}}{2}B_{12}^{x;z}C_{12}^{y;v}-\frac{\sqrt{3}}{4}(C_{12}^{y;v})^2,\\
\mathcal{T}^{(2)}&=-\frac{3}{4}(B_{12}^{x;z})^2-\sqrt{3}C_{12}^{y;u}A_{12}^{y;z}+\sqrt{3}A_{12}^{y;z}C_{12}^{y;z} \nonumber\\
&-\frac{1}{2}B_{12}^{x;z}C_{12}^{y;v}-\frac{3}{4}(C_{12}^{y;v})^2,\\
\mathcal{T}^{(3)}&=\frac{\sqrt{3}}{4}B_{12}^{x;z}A_{12}^{x;z}-\frac{1}{4}B_{12}^{x;z}C_{12}^{y;u}-\frac{1}{2}B_{12}^{x;z}C_{12}^{y;z}\nonumber\\
&+\frac{\sqrt{3}}{4}A_{12}^{x;z}C_{12}^{y;v}+\frac{3}{4}C_{12}^{y;u}C_{12}^{y;v}.
\end{align}
Here, $C_{12}^{y;u}=(C_{12}^{y;x}+C_{12}^{y;y})/2$ and $C_{12}^{y;v}=(C_{12}^{y;x}-C_{12}^{y;y})/2$.
These interactions are classified into the irreducible representation $A'_2$ under the point group $\textbf{G}=m2\bar{6}$, i.e., the totally symmetric irreducible representation under the black and white magnetic point group $\textbf{M}=m'2'\bar{6}$.

The stable spin configurations are investigated by numerically solving the LLG equation in the same manner in Sec.~\ref{sec:H2spin_tri}.
Figures~\ref{fig:3site}(a)-\ref{fig:3site}(c) show the time evolutions of $\chi_{\rm sc}$ and stable spin configurations by $\mathcal{T}^{\mathrm{(1)}}=1$, $\mathcal{T}^{\mathrm{(2)}}=1$, or $\mathcal{T}^{\mathrm{(3)}}=1$, respectively.  
We find that $\mathcal{T}^{\mathrm{(1)}}$ and $\mathcal{T}^{\mathrm{(3)}}$ favor noncoplanar spin configurations with nonzero spin scalar chirality, while $\mathcal{T}^{\mathrm{(2)}}$ favors the collinear spin configuration without the spin scalar chirality but with the uniform out-of-plane magnetization. 
Thus, the three-site three-spin interactions also favor the spin configurations belonging to $A'_2$, which is similar to the situation under the two-site three-spin interactions in Sec.~\ref{sec:HQ_tri}.
Among $\mathcal{T}$, the mechanism of nonzero $\chi_\mathrm{sc}$ is described by the coupling between these interactions and the spin scalar chirality, as clearly shown in $\mathcal{H}^{(1)}_{\rm 3site}$.  
The sign of $\chi_\mathrm{sc}$ is fixed by $\mathcal{T}^{(1)}$, while not fixed by $\mathcal{T}^{(3)}$.

\section{Summary}
\label{sec:summary}

To summarize, we have investigated the relationship between light-induced magnetic interactions and symmetry lowering by light. 
By adopting Floquet formalism, we have systematically shown the light-induced two-spin and three-spin interactions for all the crystallographic point groups in two-dimensional insulating magnets. 
In particular, the light-induced DM interaction ubiquitously appears for all the point groups, which means the possibility to stabilize helical and skyrmion states in any crystal structures.
Based on the symmetry argument, we have revealed that the emergent two-spin and three-spin interactions are the consequence of the reduction from the point group to the chiral point group and the black and white magnetic point group, respectively.   
These results have uncovered the effect of symmetry lowering by light on magnetic interactions.  
We have also shown that the light-induced magnetic interactions on the $m2m$ triangular unit favor noncoplanar spin textures with the spin scalar chirality as an example.  
Our results will give a symmetry-based understanding of controlling magnetic interactions and enable systematic Floquet engineering of magnetic structures based on crystal symmetry.

\begin{acknowledgments}
We thank S. Kitamura for the fruitful discussions.
This research was supported by JSPS KAKENHI Grants Numbers JP19K03752, JP19H01834, JP21H01037, JP22H04468, JP22H00101, JP22H01183, and by JST PRESTO (JPMJPR20L8). 
R.Y. was supported by Forefront Physics and Mathematics Program to Drive Transformation (FoPM).
\end{acknowledgments}

\clearpage
\appendix

\begin{widetext}
\section{Coupling matrix and third-rank ME tensor in the global Cartesian spin coordinate}
\label{app:matrix}

We show the coupling matrix and third-rank ME tensor in Sec.~\ref{sec:application} in the global Cartesian spin coordinate, which are given by
\begin{align}
J_{12} &= \begin{pmatrix}
F^x & D^z & 0 \\
-D^z & F^y & 0 \\
0 & 0 & F^z 
\end{pmatrix},
J_{23} = \begin{pmatrix}
\frac{1}{4}(F^x+3F^y) & \frac{\sqrt{3}}{4}(-F^x+F^y)+D^z & 0 \\
\frac{\sqrt{3}}{4}(-F^x+F^y)-D^z & \frac{1}{4}(3F^x+F^y) & 0 \\
0 & 0 & F^z 
\end{pmatrix},\\
J_{31} &= \begin{pmatrix}
\frac{1}{4}(F^x+3F^y) & \frac{\sqrt{3}}{4}(F^x-F^y)+D^z & 0 \\
\frac{\sqrt{3}}{4}(F^x-F^y)-D^z & \frac{1}{4}(3F^x+F^y) & 0 \\
0 & 0 & F^z 
\end{pmatrix},
Y^x_{12}=
\begin{pmatrix}
0 & B_{12}^{x;z} & 0 \\
B_{12}^{x;z} & 0 & 0\\
0 & 0 & 0
\end{pmatrix},Y^y_{12}=
\begin{pmatrix}
C_{12}^{y;x} & A_{12}^{y;z} & 0 \\
-A_{12}^{y;z} & C_{12}^{y;y} & 0\\
0 & 0 & C_{12}^{y;z}
\end{pmatrix},\\
Y_{23}^x&=
\begin{pmatrix}
-\frac{\sqrt{3}}{8}(C_{12}^{y;x}+3C_{12}^{y;y}+2B_{12}^{x;z}) & \frac{1}{8}(3C_{12}^{y;x}-3C_{12}^{y;y}+2B_{12}^{x;z})-\frac{\sqrt{3}}{2}A_{12}^{y;z}   & 0 \\
\frac{1}{8}(3C_{12}^{y;x}-3C_{12}^{y;y}+2B_{12}^{x;z})+\frac{\sqrt{3}}{2}A_{12}^{y;z} & -\frac{\sqrt{3}}{8}(3C_{12}^{y;x}+C_{12}^{y;y}-2B_{12}^{x;z}) & 0\\
0 & 0 & -\frac{\sqrt{3}}{2}C_{12}^{y;z}
\end{pmatrix},\\
Y_{23}^y&=
\begin{pmatrix}
-\frac{1}{8}(C_{12}^{y;x}+3C_{12}^{y;y}-6B_{12}^{x;z}) & \frac{\sqrt{3}}{8}(C_{12}^{y;x}-C_{12}^{y;y}-2B_{12}^{x;z})-\frac{1}{2}A_{12}^{y;z}   & 0 \\
\frac{\sqrt{3}}{8}(C_{12}^{y;x}-C_{12}^{y;y}-2B_{12}^{x;z})+\frac{1}{2}A_{12}^{y;z} & -\frac{1}{8}(3C_{12}^{y;x}+C_{12}^{y;y}+6B_{12}^{x;z}) & 0\\
0 & 0 & -\frac{1}{2}C_{12}^{y;z}
\end{pmatrix},\\
Y_{31}^x&=
\begin{pmatrix}
\frac{\sqrt{3}}{8}(C_{12}^{y;x}+3C_{12}^{y;y}+2B_{12}^{x;z}) & \frac{1}{8}(3C_{12}^{y;x}-3C_{12}^{y;y}+2B_{12}^{x;z})+\frac{\sqrt{3}}{2}A_{12}^{y;z}   & 0 \\
\frac{1}{8}(3C_{12}^{y;x}-3C_{12}^{y;y}+2B_{12}^{x;z})-\frac{\sqrt{3}}{2}A_{12}^{y;z} & \frac{\sqrt{3}}{8}(3C_{12}^{y;x}+C_{12}^{y;y}-2B_{12}^{x;z}) & 0\\
0 & 0 & \frac{\sqrt{3}}{2}C_{12}^{y;z}
\end{pmatrix},\\
Y_{31}^y&=
\begin{pmatrix}
-\frac{1}{8}(C_{12}^{y;x}+3C_{12}^{y;y}-6B_{12}^{x;z}) & -\frac{\sqrt{3}}{8}(C_{12}^{y;x}-C_{12}^{y;y}-2B_{12}^{x;z})-\frac{1}{2}A_{12}^{y;z}   & 0 \\
-\frac{\sqrt{3}}{8}(C_{12}^{y;x}-C_{12}^{y;y}-2B_{12}^{x;z})+\frac{1}{2}A_{12}^{y;z} & -\frac{1}{8}(3C_{12}^{y;x}+C_{12}^{y;y}+6B_{12}^{x;z}) & 0\\
0 & 0 & -\frac{1}{2}C_{12}^{y;z}
\end{pmatrix}.
\end{align}

\section{Bases of irreducible representation}
\label{app:bases}

The eigenbases $S^\alpha(\Gamma)$ in Fig.~\ref{fig:irrep} are given by 
\begin{align}
S(A'_2)&=(0,0,1,0,0,1,0,0,1), \\
S(A''_1)&=\frac{1}{2}(-\sqrt{3},-1,0,\sqrt{3},-1,0,0,2,0), \\
S(A''_2)&=\frac{1}{2}(1,-\sqrt{3},0,1,\sqrt{3},0,-2,0,0), \\
S^x(E'_{z})&=\frac{1}{\sqrt{2}}(0,0,-1,0,0,-1,0,0,2), \\
S^y(E'_{z})&=\sqrt{\frac{3}{2}}(0,0,1,0,0,-1,0,0,0), \\
S^x(E''_{\mathrm{FM}})&=(0,-1,0,0,-1,0,0,-1,0), \\
S^y(E''_{\mathrm{FM}})&=(1,0,0,1,0,0,1,0,0), \\
S^x(E''_{\mathrm{AFM}})&=\frac{1}{2}(\sqrt{3},-1,0,-\sqrt{3},-1,0,0,2,0), \\
S^y(E''_{\mathrm{AFM}})&=\frac{1}{2}(-1,-\sqrt{3},0,-1,\sqrt{3},0,2,0,0),
\end{align}
where $S^\alpha(\Gamma)\cdot S^\beta(\Gamma')=3\delta_{\Gamma\Gamma'}\delta_{\alpha\beta}$.

\end{widetext}

\bibliography{main.bbl}
\end{document}